\documentstyle[aps,preprint,epsfig]{revtex} 
\begin{document} 
\draft 
\title{Quantum-field-theoretical techniques 
for stochastic representation of quantum problems} 
\author{L.\ I.\ Plimak$^{1,2}$, M.\ Fleischhauer$^{3}$, 
M.\ K.\ Olsen$^{1}$ and M.\ J.\ Collett$^{1}$} 
\address{$^{1}$Department of Physics, University of Auckland, Private Bag 
92019, Auckland, New Zealand\\  
$^{2}$Department of Chemical Physics, The Weizmann 
Institute of Science, 76100 Rechovot, Israel\\  
$^{3}$Fachbereich Physik, Universit\"at Kaiserslautern, 
D-67663 Kaiserslautern, Germany} 
\date{\today} 
\maketitle 
\begin{abstract} 
We describe quantum-field-theoretical (QFT) techniques 
for mapping quantum problems onto c-number 
stochastic problems. 
This approach yields results which are 
identical to phase-space techniques 
[C.W. Gardiner, {\em Quantum Noise\/} 
(1991)] when the latter result in a Fokker-Planck 
equation for a corresponding pseudo-probability distribution.  
If phase-space techniques 
do not result in a Fokker-Planck equation and hence fail 
to produce a stochastic representation, 
the QFT techniques nevertheless yield stochastic difference 
equations in discretised time.
\end{abstract} 
\section{Introduction} 

There is a well-known duality between Fokker-Planck equations (FPE) 
and stochastic differential equations (SDE), which goes back as far 
as Einstein's and 
Langevin's theories of Brownian motion 
(see Risken's book \cite{Risken} for a detailed discussion of 
FP equation and related issues). Langevin equations or, more 
generally, stochastic differential equations, have long been a 
successful computational tool in quantum stochastics 
\cite{GardinerQN,DFW}, allowing for 
the numerical stochastic integration of systems for which analytical 
solution would be, at best, extremely difficult. There are 
long-standing rules, especially in quantum optics, for beginning with 
a particular system Hamiltonian and mapping this onto stochastic 
equations of motion for the field variables. However, this process 
depends on there being an FP equation equivalent to the master 
equation for the Hamiltonian in question; that is the partial 
differential equation for the appropriate probability or 
pseudoprobability distribution must contain derivatives of no higher 
than second order. This is the content of 
Pawula's theorem \cite{Risken,Pawula}. 
It 
restricts the method to a certain class of 
problems which, although containing many interesting cases, is not 
exhaustive. A generalised FP equation with higher order derivatives 
has no mapping onto a stochastic differential equation. As 
descriptions in terms 
of FP equations proper are the exception rather 
than the rule, it is of clear interest to extend present methods to 
allow for at least the numerical modelling of processes involving 
higher order noises. 

For purposes of numerical simulations, 
time may always be regarded as discretised, so 
the development of a stochastic difference equation 
{\em approximating\/} a generalised FP 
will generally be sufficient. 
Since Pawula's theorem only applies 
in the continuous time limit, 
this creates the loophole we need. 
What we will show here is that 
discretised stochastic equations may be devised for generalised 
FPEs, which, in a manner reminiscent of the 
positive-P representation of quantum optics, require a doubled 
phase-space. 
In agreement with Pawula's theorem, 
the method we use to develop these noises 
has no natural extension to the continuous time limit. 
This notwithstanding, any 
average corresponding to a physical quantity is expected to be 
correct in this limit (whereas the rest of the averages 
diverge). 
The question whether a stochastic process in a certain 
generalised mathematical sense can be defined corresponding to our 
methods is a subject for 
futher investigation. This is not, however, a problem for practical 
computer simulations, as these are necessarily performed on a 
discrete time grid. 

As in the well-known positive-P representation 
\cite{GardinerQN,DFW}, nonphysical averages in our 
method determine the sampling noise. 
The fact that they are expected to diverge in the continuous 
time limit means that sampling noise grows 
as the time-grid spacing descreases. This is 
in contrast to the properties of the Wiener process, 
where sampling noise is independent 
of the grid spacing for a given sample size. 
In practice, however, this 
distinction between the Gaussian and 
higher-order noises is quantitative rather than 
qualitative, because for a given 
computing time the sample size 
necessarily decreases with the grid spacing. 
In either case the growth of sampling noise 
prevents the grid being made too fine. 

As a specific example, in this paper we consider a 
quantum oscillator with Kerr nonlinearity. 
It is well known that the Wigner function of this 
system obeys a generalised FP 
equation with 3rd order derivatives. 
Our goal is to show that the latter is 
(approximately) equivalent to a system of stochastic 
difference equations. 
This equivalence is not straighforward but rather 
bears a lot of similarity to the positive-P 
representation known in quantum optics 
\cite{GardinerQN,DFW}. 
We shall therefore call these equations 
a positive-W representation of the nonlinear 
quantum oscillator. 

The positive-P representation 
originates with the single-time P-distribution, 
yet it allows one to 
calculate a much wider class of quantum 
averages \cite{GardinerQN,DFW}, namely, averages 
of multi-time, time-normally 
ordered \cite{Glauber} operator products. 
The positive-P representation may thus be regarded as 
a constructive mapping of a quantum problem to a classical 
stochastic problem. 
Its characterictic property is that time-normal averages 
are mapped directly on classical averages. 
In a much similar manner, 
the positive-W equations, which originate 
with the single-time W-distribution, 
allow one to calculate multi-time, {\em time-Wigner ordered\/} 
operator averages. 
This new type of operator ordering is introduced in 
this paper; to the best of our knowledge 
it has never been discussed before. 
It generalises the 
single-time symmetric operator ordering, 
in the same way as the time-normal ordering generalises the 
single-time normal ordering. 
The positive-W representation is hence 
a constructive mapping of a quantum problem onto a classical 
stochastic problem, with the characterictic property 
of relating time-Wigner averages 
directly to classical averages. 

The question then is if, 
rather than taking the usual route from q-number 
to c-number equations based on phase-space 
techniques 
\cite{GardinerQN,DFW}, 
the positive-P 
and the positive-W representations 
may be derived by directly 
linking q-number Heisenberg 
equations to c-number Langevin equations. 
In this paper, we show that such a derivation is indeed 
possible, by employing methods \cite{Corr,EPL} 
based on the techniques of quantum field theory. 
Our considerations 
closely follow the way in which Feynman 
diagram techniques were derived in textbooks 
dating back to the fifties and sixties \cite{QFT}. 
We derive Keldysh diagram series 
\cite{Keldysh} for the Kerr oscillator, 
then recast them 
as a Wyld-type series \cite{Wyld}, 
otherwise termed {\em causal\/} series \cite{EPL,ClDiags} 
(see also Ref.\ \cite{SqOPO}). 
(More precisely speaking, we derive generating expressions 
\cite{ClDiags} for these series, avoiding 
diagram notation as such.) 
Causal series emerge as solutions to 
c-number stochastic problems \cite{ClDiags}. 
On the other hand, given a causal 
series, it is easy to write 
a stochastic differential (or difference) equation 
for which this series is a solution \cite{ClDiags}. 
This yields strikingly 
simple and powerful techniques for 
obtaining stochastic representations of quantum problems. 
(Cf., e.g., section \ref{GenOPO} 
where positive-P equations are derived for 
an optical parametric oscillator, 
the essence of the derivation being a two-line calculation.) 
Not the least important property of these techniques is 
that they can be formulated using simple recipes and then used 
without any reference to the advanced methods employed 
in their derivation. 

The paper is structured as follows. 
In section \ref{SecQFT}, we reiterate the standard 
techniques of quantum field theory as applied to a 
nonlinear oscillator. We discuss in detail causal 
regularisation \cite{ClDiags} of the propagator, 
which is necessary to order to make 
our relations unambiguous. 
A functional (Hori's) form of Wick's theorem 
\cite{Hori} is introduced 
for the normal ordering and then generalised to the 
case of symmetric ordering. 
The result of the section is a pair of closed perturbative 
relations. 
One of them applies when the inital state of the field 
is characterised via a P-distribution; 
the other applies if this state is characterised 
via a W-distribution \cite{GardinerQN,DFW}. 
In section \ref{SecCaus}, we investigate the 
causal structure of the perturbative relations derived in 
section \ref{SecQFT}. 
The concept of causality is introduced via the 
retarded Green's function of the free Schr\"odinger equation. 
Following it, we are able to define the {\em input and output\/} of 
a quantum system. 
We then show that, physically, the input and output 
thus introduced correspond to 
a generalisation of Kubo's linear reaction approach 
\cite{Kubo} to 
a full nonlinear quantum-stochastic response problem. 
In section \ref{SecStoch}, we develop techniques (based 
on the Hubbard-Stratonovich transformation \cite{HST}) which 
allow for a constructive mapping of the quantum nonlinear 
response 
problem onto a classical nonlinear stochastic response problem. 
For the oscillator, this results in the positive-P 
and positive-W representations which we are seeking. 
We complete the section with results of 
computer simulations of the Kerr oscillator using these 
representations. 
Finally, in section \ref{SecCook} we reformulate our 
results recipe-style, the way they should be applied 
in calculations. 
Despite all the tediousness of their derivation, 
we end up with two simple relations, which 
allow one to derive positive-P and positive-W 
representations 
for an arbitrary nonlinear quantum system. 
We illustrate their use by applying them to 
a number of systems 
common in quantum optics. 
\section{Quantum field theory of the Kerr oscillator} 
\label{SecQFT} 
\subsection{The model} 
The techniques we introduce in this paper are 
applicable to any Hamiltonians which have 
a polynomial form in the field operators. 
We also assume that the Hamiltonian can be divided into a quadratic 
part, called the free Hamiltonian, and the 
remainder, termed the interaction Hamiltonian. 
This allows us to introduce, in the usual manner, 
Schr\"odinger, Heisenberg, and interaction-picture field 
operators. 
However, this requirement may always be satisfied by 
subtracting a suitable quadratic term from the Hamiltonian, 
and declaring the remainder as being the ``interaction'' part 
(cf.\ the way a Higgs-type phase transition in an anharmonic 
oscillator was treated in Ref.\ \cite{EPL}). 

This generality notwithstanding, our techniques may be effectively 
demonstrated for a 1D oscillator 
(as is usual in quantum 
field theory). 
We therefore consider a 
nonlinear quantum oscillator with the Hamiltonian, 
\begin{eqnarray} 
\label{eq:H} 
{\cal H} &=& 
{\cal H}_0 + {\cal H}_{\text{int}} = 
\omega \hat a_{\text{S}} 
^{\dag} 
\hat a_{\text{S}} + 
\frac{\kappa }{4}\, \hat a_{\text{S}} 
^{\dag 2}%
\hat a_{\text{S}} ^2 
, 
\end{eqnarray}%
using units such that $\hbar =1$. 
In (\ref{eq:H}), $\hat a_{\text{S}} 
^{\dag}%
$ and $\hat a_{\text{S}} $ are 
the usual pair of creation and 
annihilation operators with commutator $[\hat a%
_{\text{S}} 
,\hat a%
_{\text{S}} 
^{\dag}%
]=1$. 
They play the role of Schr\"odinger-picture field 
operators for this illustrative system. 
The field operators in the interaction picture 
are simply 
\begin{eqnarray} 
\label{eq:at} 
\hat{a}(t) = 
\text{e}^{%
-i\omega t%
}%
\hat{a}%
_{\text{S}}%
, \ \ \ 
\hat{a}%
^{\dag}%
(t) = 
\text{e}^{%
i\omega t%
}%
\hat{a}%
_{\text{S}}%
^{\dag}%
, 
\end{eqnarray}%
while Heisenberg picture operators 
will be denoted in a slanted font as 
$\hat{\sl a}%
^{\dag}%
(t)$ and $\hat{\sl a}(t)$. 
\subsection{Time orderings of operators} 
Time ordering of operators 
puts operators from right to left in the order 
of {\em increasing\/} time arguments, e.g., 
\begin{eqnarray} 
\label{eq:Tp} 
T_+ \hat{\sl a}(t') \hat{\sl a}%
^{\dag}%
(t) = 
\left \{ 
\begin{array}{cc} 
\hat{\sl a}%
^{\dag}%
(t)\hat{\sl a}(t'), & t\geq t',\\ 
\hat{\sl a}(t')\hat{\sl a}%
^{\dag}%
(t), & t< t' . 
\end{array} 
\right . 
\end{eqnarray}%
For equal times, the time ordering is 
specified as normal ordering 
(which places all creation operators 
on the left of annihilation operators). 
Further, the reverse time ordering, 
$T_-$, places operators in the order 
of {\em decreasing\/} time arguments. 
Formally, it may be defined as a conjugate of $T_+$: 
\begin{eqnarray} 
T_- \hat P = 
\left [ 
T_+ 
\left ( 
\hat P%
^{\dag} 
\right ) 
\right ]%
^{\dag} 
, 
\end{eqnarray}%
where $\hat P$ is a product of field operators. 
Then, e.g., 
\begin{eqnarray} 
T_- \hat {\sl a}(t') \hat{\sl a}%
^{\dag}%
(t) = 
\left \{ 
\begin{array}{cc} 
\hat{\sl a}%
^{\dag}%
(t)\hat{\sl a}(t'), & t\leq t',\\ 
\hat{\sl a}(t')\hat{\sl a}%
^{\dag}%
(t), & t> t' . 
\end{array} 
\right . 
\end{eqnarray}%
For equal times, $T_-$ also becomes 
normal ordering. 
Double time ordering is 
the combination of the $T_+$ and $T_-$-orderings, 
\begin{eqnarray} 
\label{eq:TT} 
T_- \hat P_- \ 
T_+ \hat P_+ , 
\end{eqnarray}%
where $\hat P_-$ and $\hat P_+$ are operator products 
(by definition, 
$T_-$ acts on $\hat P_-$ and $T_+ $ acts on $\hat P_+$). 
\subsection{Non-stationary perturbation approach} 
\subsubsection{Perturbation expressions 
for time-ordered operator products} 
Heisenberg operators are related to those 
in the interaction picture via the evolution 
operator, 
\begin{eqnarray} 
\hat{\sl a}(t) &=& 
{\cal U}%
^{\dag}%
(t,-\infty)\hat{a}(t){\cal U}(t,-\infty), 
\\ 
\hat{\sl a}%
^{\dag}%
(t) &=& 
{\cal U}%
^{\dag}%
(t,-\infty)\hat{a}%
^{\dag}%
(t){\cal U}(t,-\infty) , 
\end{eqnarray}%
which is a solution to 
the Schr\"odinger equation, 
\begin{eqnarray} 
i\frac{\partial {\cal U}(t,t_0)} 
{\partial t} = {\cal H}_{\text{int}} (t) 
{\cal U}(t,t_0), \ \ \ 
{\cal U}(t_0,t_0) = \openone , 
\end{eqnarray}%
where 
\begin{eqnarray} 
\label{eq:Hint} 
{\cal H}_{\text{int}} (t) 
= 
\frac{\kappa }{4}\, \hat a 
^{\dag 2} 
(t)\hat a ^2 (t) 
. 
\end{eqnarray}%
The evolution operator may be written as a 
time-ordered operator expression (T-exponent): 
\begin{eqnarray} 
\label{eq:Ev} 
{\cal U}(t,t_0) = T_+ \exp 
\left [ 
-i\int_{t_0}^t dt' {\cal H}_{\text{int}} (t') 
\right ]. 
\end{eqnarray}%
It is unitary, and has the group property, 
\begin{eqnarray} 
{\cal U}(t,t'){\cal U}(t',t'') = 
{\cal U}(t,t''), \ \ \ t\geq t' \geq t'' . 
\end{eqnarray}%
In particular it follows that 
\begin{eqnarray} 
{\cal U}(t,t') = 
{\cal U}(t,-\infty) 
{\cal U}%
^{\dag}%
(t',-\infty). 
\end{eqnarray}%
Then, assuming that times are ordered in a certain way, $t>t'$, 
we have: 
\begin{eqnarray} 
T_+ \hat{\sl a}%
^{\dag}%
(t)\hat{\sl a}(t') = 
{\cal U}%
^{\dag}%
(\infty,-\infty) 
{\cal U}(\infty,t) 
\hat{a}%
^{\dag}%
(t) 
{\cal U}(t,t') 
\hat{a}(t') 
{\cal U}(t',-\infty) . 
\end{eqnarray}%
It should be noted here that the combination 
\begin{eqnarray} 
{\cal U}(\infty,t) 
\hat{a}%
^{\dag}%
(t) 
{\cal U}(t,t') 
\hat{a}(t') 
{\cal U}(t',-\infty) 
\end{eqnarray}%
is time-ordered. 
Finally, introducing the S-matrix, 
\begin{eqnarray} 
{\cal S} = {\cal U}(\infty,-\infty), 
\end{eqnarray}%
we have, 
\begin{eqnarray} 
T_+ \hat{\sl a}%
^{\dag}%
(t)\hat{\sl a}(t') 
&=& 
{\cal S}%
^{\dag} 
T_+ \hat{a}(t')\hat{a}%
^{\dag} 
(t) 
\exp 
\left [ - \frac{i \kappa }{4} 
\int_{-\infty}^{+\infty}d\tau 
\hat a%
^{\dag 2}%
(\tau)\hat a^{2}(\tau) 
\right ] 
\label{eq:Taa} 
. 
\end{eqnarray}%

There is a certain subtlety in the way in which relations such as 
(\ref{eq:Ev}) and (\ref{eq:Taa}) should be understood. 
Quantities under ordering 
should be regarded as functions 
of $\hat a(t)$ and $\hat a%
^{\dag}%
(t)$, 
whereas (\ref{eq:at}) may be used only after 
the time ordering has been completed. 
So calculating the time integral in (\ref{eq:Taa}) 
should follow the ordering rather than precede it. 
Were one to take the integral before 
the time ordering, 
it would prevent $T_+$ from acting on the operators 
comprising the interaction Hamiltonian. 

\subsubsection{Functional perturbation techniques} 
For our purposes we need relations applying to 
the whole assemblage of time-ordered operator products. 
An arbitrary 
time-ordered operator product 
may be found by differentiating the 
following time-ordered operator exponent, 
\begin{eqnarray} 
\label{eq:TpF} 
T_+ \exp 
\left \{ 
\int_{-\infty}^{+\infty} dt 
\left [ 
\zeta (t) \hat {\sl a}%
^{\dag} 
(t) 
+ 
\zeta 
^{\dag}%
(t) \hat {\sl a} (t) 
\right ] 
\right \} 
\equiv T_+ \exp 
\left ( 
\zeta \hat {\sl a}%
^{\dag} 
+ 
\zeta 
^{\dag} 
\hat {\sl a} 
\right ) 
, 
\end{eqnarray}%
where $\zeta (t), \zeta%
^{\dag} 
(t) $ is a pair of 
c-number functions. Then, e.g., 
\begin{eqnarray} 
T_+ \hat{\sl a}(t')\hat{\sl a}%
^{\dag}%
(t) 
= 
\frac{\delta ^2}{\delta \zeta (t)\delta \zeta%
^{\dag} 
(t')} 
\, T_+ \exp 
\left ( 
\zeta \hat {\sl a}%
^{\dag} 
+ 
\zeta 
^{\dag} 
\hat {\sl a} 
\right ) 
|_{\zeta =\zeta 
^{\dag} 
= 0}, 
\end{eqnarray}%
and so on. Equation (\ref{eq:Taa}) is readily 
generalised, resulting in 
\begin{eqnarray} 
\nonumber 
T_+ \exp 
\left ( 
\zeta \hat {\sl a}%
^{\dag} 
+ 
\zeta 
^{\dag} 
\hat {\sl a} 
\right ) 
&=& 
{\cal S}%
^{\dag} 
T_+ \exp 
\left \{ 
\int_{-\infty}^{+\infty} dt 
\left [ 
\zeta (t) \hat { a}%
^{\dag} 
(t) 
+ 
\zeta 
^{\dag}%
(t) \hat { a} (t) 
- \frac{i \kappa }{4}\, 
\hat a%
^{\dag 2}%
(t)\hat a^{2}(t) 
\right ] 
\right \} 
\\ 
&\equiv & 
{\cal S}%
^{\dag} 
T_+ \exp 
\left ( 
\zeta \hat { a}%
^{\dag} 
+ 
\zeta 
^{\dag} 
\hat { a} 
- \frac{i \kappa }{4}\, 
\hat a%
^{\dag 2}%
\hat a^{2} 
\right ) 
. 
\label{eq:%
TFFull%
} 
\end{eqnarray}%
Relations for the inverse and double time 
orderings then read 
\begin{eqnarray} 
&&T_- \exp 
\left ( 
\zeta \hat {\sl a}%
^{\dag} 
+ 
\zeta 
^{\dag} 
\hat {\sl a} 
\right ) 
= 
T_- \exp 
\left ( 
\zeta \hat { a}%
^{\dag} 
+ 
\zeta 
^{\dag} 
\hat { a} 
+ \frac{i \kappa }{4}\, 
\hat a%
^{\dag 2}%
\hat a^{2} 
\right ){\cal S}, 
\\ \nonumber 
&&T_- \exp 
\left ( 
\zeta _- \hat {\sl a}%
^{\dag} 
+ 
\zeta_- 
^{\dag} 
\hat {\sl a} 
\right ) 
\, T_+ \exp 
\left ( 
\zeta _+ \hat {\sl a}%
^{\dag} 
+ 
\zeta _+%
^{\dag} 
\hat {\sl a} 
\right ) 
\\ && \hspace{5ex} 
= 
T_- \exp 
\left ( 
\zeta _- \hat { a}%
^{\dag} 
+ 
\zeta _-%
^{\dag} 
\hat { a} 
+ \frac{i \kappa }{4}\, 
\hat a%
^{\dag 2}%
\hat a^{2} 
\right ) 
\, T_+ \exp 
\left ( 
\zeta _+ \hat { a}%
^{\dag} 
+ 
\zeta _+%
^{\dag} 
\hat { a} 
- \frac{i \kappa }{4}\, 
\hat a%
^{\dag 2}%
\hat a^{2} 
\right ) . 
\label{eq:DTP} 
\end{eqnarray}%
The relation for the double time ordering (naturally) 
involves four arbitrary c-number functions, 
$\zeta_{\pm} (t), \zeta_{\pm}%
^{\dag} 
(t) $. 
Note that, in this relation, the 
factors ${\cal S}$ and ${\cal S}%
^{\dag}%
$ outside 
the orderings have cancelled 
each other. 
This results in a genuine double-time-ordered 
structure on the RHS of (\ref{eq:DTP}). 

In what follows, we will make wide use of the brief notation as in 
Eqs.\ (\protect\ref{eq:TpF})%
, (\ref{eq:TFFull}) 
and (\ref{eq:DTP}), 
which will always be signalled by the absence 
of time arguments of the fields. 
Apart from these cases, we will not omit time arguments where 
they apply, as in 
Eq.\ (\protect\ref{eq:Hint})%
, 
so as to remove any ambuguity. 
(Note that this is also the reason why the Scr\"odinger-picture field 
operators were denoted $\hat a%
_{\text{S}}%
,\hat a%
_{\text{S}}%
^{\dag}%
$, 
not $\hat a, \hat a%
^{\dag}%
$.) 
\subsection{Wick's theorems} 
\subsubsection{Hori's form of Wick's theorem proper} 
A common way of dealing with time-ordered 
expressions is by converting them 
to normally ordered form. 
This is done following Wick's theorem, 
which states that 
``a time ordered product 
of interaction-picture field operators equals the 
sum of all possible normally ordered operator products, 
obtained by 
replacing pairs of operators in the initial product by 
corresponding contractions (including the term without 
contractions)''. 
For the oscillator, we have only one nonzero contraction, namely, 
\begin{eqnarray} 
T_+\hat a(t')\hat a%
^{\dag}%
(t) - 
:\hat a%
^{\dag}%
(t)\hat a(t'):\, 
&=& 
\left\langle 0\left|%
T_+\hat a%
^{\dag}%
(t)\hat a(t')%
\right|0\right\rangle 
\nonumber \\ 
&=& \theta(t'-t)%
\text{e}^{%
-i\omega (t'-t)%
} 
\equiv i G(t'-t) . 
\end{eqnarray}%
In the above, $G(t)$ is a 
retarded Green's function of 
the free Schr\"odinger equation: 
\begin{eqnarray} 
\left ( 
i\frac{\partial}{\partial t} - \omega 
\right )G(t) 
= \delta (t) . 
\end{eqnarray}%
It then follows that 
\begin{eqnarray} 
T_+\hat a%
^{\dag}%
(t)\hat a(t') &=& 
\hat a%
^{\dag}%
(t)\hat a(t') + iG(t'-t), 
\\ 
T_+\hat a%
^{\dag}%
(t)\hat a(t') \hat a(t'') &=& 
\hat a%
^{\dag}%
(t)\hat a(t') \hat a(t'') 
+ iG(t'-t)\hat a(t'') 
+ iG(t''-t)\hat a(t') 
, 
\\ 
T_+\hat a%
^{\dag}%
(t)\hat a%
^{\dag}%
(t') 
\hat a(t'') \hat a(t''') &=& 
\hat a%
^{\dag}%
(t)\hat a%
^{\dag}%
(t') 
\hat a(t'') \hat a(t''') 
\nonumber \\ && 
+\, iG(t''-t)\hat a%
^{\dag}%
(t')\hat a(t''') + 
iG(t'''-t)\hat a%
^{\dag}%
(t')\hat a(t'') + 
\nonumber \\ && 
+\, iG(t''-t')\hat a%
^{\dag}%
(t)\hat a(t''') + 
iG(t'''-t')\hat a%
^{\dag}%
(t)\hat a(t'') 
\nonumber \\ && 
-\, G(t''-t) G(t'''-t') 
- G(t''-t') G(t'''-t) 
, 
\end{eqnarray}%
and so on. 

As was noticed by Hori~\cite{Hori}, 
the pattern of products with contractions 
is exactly that produced by a quadratic 
form of functional derivatives, 
\begin{eqnarray} 
\Delta = \int dt dt' iG(t'-t) 
\frac{\delta ^2}{\delta a(t')\delta a%
^{\dag}%
(t)} , 
\end{eqnarray}%
acting repeatedly on products of 
{\em c-number\/} functions $a(t),\ a%
^{\dag}%
(t)$. 
Applying $\Delta $ once produces terms 
with a single contraction: 
\begin{eqnarray} 
\Delta a%
^{\dag}%
(t) a(t') &=& iG(t'-t), 
\\ 
\Delta a%
^{\dag}%
(t) a(t') a(t'') &=& 
iG(t'-t) a(t'') 
+ iG(t''-t) a(t') 
, 
\\ 
\Delta a%
^{\dag}%
(t) a%
^{\dag}%
(t')  
a(t'') a(t''') &=& 
iG(t''-t) a%
^{\dag}%
(t') a(t''') + 
iG(t'''-t) a%
^{\dag}%
(t') a(t'') + 
\nonumber \\ && 
iG(t''-t') a%
^{\dag}%
(t) a(t''') + 
iG(t'''-t') a%
^{\dag}%
(t) a(t'') , 
\end{eqnarray}%
etc. 
Terms with $n$ contractions are produced by 
$\Delta ^n/n!$. As an example, consider the case where $n=2$: 
\begin{eqnarray} 
\frac{\Delta^2}{2}\, a%
^{\dag}%
(t) a%
^{\dag}%
(t')  
a(t'') a(t''') &=& 
- \, G(t''-t) G(t'''-t') 
- G(t''-t') G(t'''-t) 
. 
\end{eqnarray}%
The whole assemblage of terms 
required by Wick's theorem 
(including those without contractions) is thus found by applying 
a differential operator, 
\begin{eqnarray} 
1 + \Delta + \frac{\Delta ^2}{2} + 
\cdots = 
\text{e}^{%
\Delta 
}%
. 
\end{eqnarray}%
This allows one to write Wick's theorem in a compact form as 
(with $\hat P$ being an arbitrary operator product) 
\begin{eqnarray} 
\label{eq:%
WickTF%
} 
T_+ \hat P = \ 
: 
\left [ 
\text{e}^{%
\Delta 
} 
\left ( 
\hat P|_{ 
\hat a(t)\rightarrow a(t), 
\hat a%
^{\dag}%
(t)\rightarrow a%
^{\dag}%
(t) 
} 
\right ) 
\right ]|_{ 
a(t)\rightarrow \hat a(t), 
a%
^{\dag}%
(t)\rightarrow \hat a%
^{\dag}%
(t) 
} 
: 
\ . 
\end{eqnarray}%
\subsubsection{Causal regularisation} 
Wick's theorem requires that no contractions 
should occur between operators with equal time arguments. 
(This is simply because the time ordering was specified 
for equal times as normal ordering.) 
However, applying (\ref{eq:WickTF}) results in ambiguities, e.g., 
\begin{eqnarray} 
\Delta a%
^{\dag}%
(t)a(t) = iG(0) = i\theta(0) \text{ (undefined)}. 
\end{eqnarray}%
A convenient way around this problem is 
a {\em causal regularisation\/} of $G(t)$. To this end, 
we shall always assume that $G(t)$ is somehow smoothed 
while still preserving its causal nature: 
\mbox{$%
G(t)=0%
$} 
for 
\mbox{$t<0$}%
. 
For a continuous function this also means that 
\mbox{$%
G(0)=0%
$}%
. 
With this specification (and implying 
a final limit of unsmoothed $G$), 
Eq.\ (\protect\ref{eq:%
WickTF%
}) 
becomes fully equivalent to Wick's theorem. 

\subsubsection{Wick's theorem for double time ordering} 
It may be checked that proof of Wick's theorem \cite{QFT} 
is based only on the linear ordering of the time axis; 
consequently Wick's theorem may be generalised 
to operators defined formally on any linearly ordered set. 
This clearly applies to the double time ordering, 
Eq.\ (\protect\ref{eq:TT})%
, 
which may alternatively be 
introduced as an ordering 
on the so-called C-contour \cite{Keldysh}. 
The C-contour (see 
Fig.\ \protect\ref{figC}%
) first travels 
from 
\mbox{$%
t = - \infty%
$} 
to 
\mbox{$%
t = + \infty%
$} 
(direct branch) and 
then back to 
\mbox{$%
t = - \infty%
$} 
(reverse branch). Then, 
\begin{eqnarray} 
T_- \hat P_- 
\ 
T_+ \hat P_+ = 
T_C \hat P_- 
|_{ 
\hat {\sl a}\to\hat {\sl a}_-, 
\hat {\sl a}%
^{\dag}%
\to\hat {\sl a}_-%
^{\dag} 
} 
\hat P_+ 
|_{ 
\hat {\sl a}\to\hat {\sl a}_+, 
\hat {\sl a}%
^{\dag}%
\to\hat {\sl a}_+%
^{\dag} 
} 
, 
\label{eq:TC} 
\end{eqnarray}%
with the subscripts `$+$' and `$-$' specifying operators as 
being defined on the direct and reverse branches of the C-contour. 
(In other words, operators acquire an additional argument; 
a C-contour index. 
This argument 
is useful purely for labelling purposes and should be disregarded 
after the ordering has been performed.) 
For the $T_C$ ordering, operator contraction becomes 
a matrix in respect of the C-countour indices, (%
\mbox{$%
\alpha ,\beta = +,-%
$}%
) 
\begin{eqnarray} 
\label{eq:Gab} 
iG_{\alpha \beta }(t'-t) = 
\left\langle 0\left| 
T_C \hat a_{\alpha}(t') \hat a%
^{\dag}%
_{\beta }(t) 
\right|0\right\rangle 
, 
\end{eqnarray}%
and the functional (Hori's) form of Wick's theorem 
generalised to the double time ordering is found to be: 
\begin{eqnarray} 
\nonumber 
&&T_- 
\hat P_- 
\ 
T_+ 
\hat P_+ 
= 
\ : 
\left [ 
\text{e}^{%
\Delta_C 
} 
\left ( 
\hat P_-|_{ 
\hat a(t)\rightarrow a_-(t), 
\hat a%
^{\dag}%
(t)\rightarrow a_-%
^{\dag}%
(t) 
} 
\right . \right . \\ &&\hspace{5ex} \times \left . \left . 
\hat P_+|_{ 
\hat a(t)\rightarrow a_+(t), 
\hat a%
^{\dag}%
(t)\rightarrow a_+%
^{\dag}%
(t) 
} 
\right ) 
\right ]|_{ 
a_-(t),a_+(t)\rightarrow \hat a(t), 
a_-%
^{\dag}%
(t),a_+%
^{\dag}%
(t)\rightarrow \hat a%
^{\dag}%
(t) 
} 
: 
\ . 
\label{eq:%
WickTFD%
} 
\end{eqnarray}%
Here, $\hat P_-$ and $\hat P_+$ are arbitrary operator products, 
$a_{\pm}(t),a_{\pm}%
^{\dag}%
(t)$ are four independent c-number functions, 
and $\Delta _C$ is a quadratic form of functional derivatives, 
\begin{eqnarray} 
\nonumber 
&&\Delta_C = \int dt dt' 
\left [ 
iG_{++}(t'-t)\frac{\delta ^2}{\delta a_+(t')\delta a_+%
^{\dag}%
(t)} 
\right . \\ &&\hspace{5ex} \left . + \ 
iG_{--}(t'-t)\frac{\delta ^2}{\delta a_-(t')\delta a_-%
^{\dag}%
(t)} + 
iG_{-+}(t'-t)\frac{\delta ^2}{\delta a_-(t')\delta a_+%
^{\dag}%
(t)} 
\right ] 
, 
\label{eq:%
DeltaC%
} 
\end{eqnarray}%
where the kernels are the three nonzero components of $G_{\alpha \beta }$ (shown schematically in 
Fig.\ \protect\ref{figC}%
). 
In turn, these are conveniently expressed in terms of $G(t)$: 
\begin{eqnarray} 
\label{eq:%
GMatrix%
} 
G_{++}(t) = G(t), \ 
G_{--}(t) = -\,G^*(-t), \ 
G_{-+}(t) = G(t)- G^*(-t) 
. 
\end{eqnarray}%
Note that 
Eqs.\ (\protect\ref{eq:%
WickTFD%
}) 
and (\ref{eq:GMatrix}) 
are formulated so as to 
eliminate the concept 
of the C-contour from our considerations from here on. 

Relations (\ref{eq:GMatrix}) require a word of caution. 
For equal times, both the $T_+$ and the $T_-$ orderings are 
specified as normal ordering, 
hence the no-contractions-between-the-%
same-time-operators caveat of Wick's theorem 
applies equally to operators under the $T_-$ ordering. 
In (\ref{eq:WickTFD}), it is enforced through the equation 
relating $G_{--}(t)$ to $G(t)$ and 
the (implied) causal regularisation of $G(t)$. 
At the same time, through the equation relating $G_{-+}(t)$ to $G(t)$, 
the regularisation also modifies $G_{-+}(t)$, 
``burning a hole'' in it in the vicinity of 
\mbox{$t=0$}%
. 
Both the causal regularisation and relations (\ref{eq:GMatrix}) 
play central roles in 
our considerations, so we have to make sure that 
the stated modification of $G_{-+}(t)$ does not lead to incorrect results. 
The fact that weird results may indeed follow demonstrates, e.g., 
a ``proof'' that $\hat a%
_{\text{S}}%
$ and $\hat a%
_{\text{S}} 
^{\dag}%
$ commute. 
With regularisation, 
\mbox{$%
G_{-+}(0)=0%
$}%
, and 
we ``obtain'', 
\begin{eqnarray} 
\hat a%
_{\text{S}} 
\hat a%
_{\text{S}} 
^{\dag} 
= 
T_C\hat a_-(t) \hat a_+ 
^{\dag}%
(t) = 
\hat a%
^{\dag}%
(t) \hat a(t) + i G_{-+} (0) = 
\hat a%
_{\text{S}} 
^{\dag} 
\hat a%
_{\text{S}} 
. 
\end{eqnarray}%
The flaw in this ``proof'' is that 
all quantities we deal with 
should be regarded as generalised functions (distributions) 
and not pointwise functions. 
This is especially true under regularisation. 
``Holes'' in continuous 
functions which emerge due to 
regularisation should be simply ignored (smeared out). 
We would only expect problems associated 
with the ``holes'' if they were overlapping with 
sufficiently strong singularities, $\delta $-functions or worse, 
whereas the worst type of singularity that 
we may expect to occur is a step-function. 
The ``hole'' in $G_{-+}(t)$ should thus be of no consequence. 

\subsubsection{Reordering time-ordered operators symmetrically} 
Wick's theorem also holds if the desired 
result is symmetric, or Wigner, operator ordering. 
(For a definition of the symmetric ordering see, 
e.g. \cite{Agarwal}.) 
The easiest way to demonstrate this is to derive an analog 
of 
Eq.\ (\protect\ref{eq:%
WickTFD%
})%
. 
A relation between a normal and a symmetric representation 
of a particular Scr\"odinger operator, 
most suitable for our purposes, 
is given in Ref.\ \cite{Agarwal}. 
It can be readily adapted to operators in the 
interaction picture resulting in 
\begin{eqnarray} 
\label{eq:N2S} 
: \hat P : \ = 
\text{W} 
\left \{ 
\left [ 
\exp 
\left ( i 
\frac{\delta }{\delta a} 
G_W 
\frac{\delta }{\delta a%
^{\dag}%
} 
\right ) 
\left ( 
\hat P|_{\hat a(t)\to a(t),\hat a%
^{\dag}%
(t)\to a%
^{\dag}%
(t)} 
\right ) 
\right ] 
|_{a(t)\to \hat a(t),a%
^{\dag}%
(t)\to \hat a%
^{\dag}%
(t)} 
\right \} 
. 
\end{eqnarray}%
Here, $\hat P$ is an arbitrary operator product, 
W denotes Wigner ordering, and 
\begin{eqnarray} 
\label{eq:N2S2} 
\frac{\delta }{\delta a} 
G_W 
\frac{\delta }{\delta a%
^{\dag}%
} = \int dt dt' 
G_W(t'-t) 
\frac{\delta^2 }{\delta a(t')\delta a%
^{\dag}%
(t)} 
\ , 
\end{eqnarray}%
where the kernel $G_W$ is defined as 
\begin{eqnarray} 
i G_W(t-t') =  
\hat a%
^{\dag}%
(t)\hat a(t') - 
\text{W} 
\hat a(t') a%
^{\dag}%
(t) =  
- 
\left\langle 0\left|%
\text{W} 
\hat a(t') a%
^{\dag}%
(t)%
\right|0\right\rangle 
=  
-\,\frac{\displaystyle 
\text{e}^{%
-i\omega (t'-t)%
}%
}{2} 
. 
\end{eqnarray}%
Note that the type of 
brief notation introduced by (\ref{eq:N2S2}) 
accounts for asymmetry of the kernel $G_W$. 
We now make use of the following fancy form of the 
rule of product differentiation, 
\begin{eqnarray} 
\label{eq:PDiff} 
\Phi 
\left ( 
\frac{\delta }{\delta \varphi } 
\right ) 
\Phi _1 
\left ( 
\varphi 
\right ) 
\Phi _2 
\left ( 
\varphi 
\right ) = 
\Phi 
\left ( 
\frac{\delta }{\delta \varphi _1} 
+ 
\frac{\delta }{\delta \varphi_2 } 
\right ) 
\Phi _1 
\left ( 
\varphi_1 
\right ) 
\Phi _2 
\left ( 
\varphi_2 
\right ) 
|_{ 
\varphi_1,\varphi_2=\varphi 
} 
. 
\end{eqnarray}%
where 
$\varphi (t), \ \varphi_1 (t), \ \varphi_2(t)$ 
are c-number functions and 
$\Phi(\varphi ) , \ \Phi_1(\varphi ) , \ \Phi_2(\varphi )$ 
are functionals 
of such functions. 
Combining (\ref{eq:WickTFD}) and (\ref{eq:N2S}) 
and employing (\ref{eq:PDiff}), 
we arrive at a relation, 
\begin{eqnarray} 
\nonumber 
&&T_- 
\hat P_- 
\ 
T_+ 
\hat P_+ 
= 
\text{W}%
\left\{ 
\left [ 
\text{e}^{%
\tilde \Delta_C^W 
} 
\left ( 
\hat P_-|_{ 
\hat a(t)\rightarrow a_-(t), 
\hat a%
^{\dag}%
(t)\rightarrow a_-%
^{\dag}%
(t) 
} 
\right . \right . \right . 
\\ &&\hspace{5ex} \times \left . \left . \left . 
\hat P_+|_{ 
\hat a(t)\rightarrow a_+(t), 
\hat a%
^{\dag}%
(t)\rightarrow a_+%
^{\dag}%
(t) 
} 
\right ) 
\right ]|_{ 
a_-(t),a_+(t)\rightarrow \hat a(t), 
a_-%
^{\dag}%
(t),a_+%
^{\dag}%
(t)\rightarrow \hat a%
^{\dag}%
(t) 
} \right\} 
, 
\label{eq:%
WickTFDWI%
} 
\end{eqnarray}%
where 
\begin{eqnarray} 
\label{eq:%
DTilde%
} 
\tilde\Delta _C^W 
= 
i\sum_{\alpha ,\beta = +,-} 
\frac{\delta }{\delta a_{\alpha }} 
\left ( 
G_{\alpha \beta} +G_W 
\right ) 
\frac{\delta }{\delta a_{\beta }%
^{\dag}%
} 
. 
\end{eqnarray}%

For nonzero time arguments, kernels comprising (\ref{eq:DTilde}) 
may be expressed by the retarded Green's function, $G(t)$. 
We note that 
\begin{eqnarray} 
G_{\alpha \beta}(t) +G_W(t) = 
G_{\alpha \beta}(t) - \frac{1}{2}[G(t) - G^*(-t)] 
\equiv G^W_{\alpha \beta}(t). 
\end{eqnarray}%
Then, employing (\ref{eq:GMatrix}), we find, 
\begin{eqnarray} 
\label{eq:%
GWMatrix%
} 
G^W_{++}(t) = - G^W_{--}(t) = \frac{1}{2} 
\left [ 
G(t) + G^*(-t) 
\right ] 
, \ \ 
G^W_{-+}(t) = - G^W_{+-}(t) = \frac{1}{2} 
\left [ 
G(t) - G^*(-t) 
\right ] 
. 
\end{eqnarray}%
Extending these relations to all times, we {\em define\/}, 
\begin{eqnarray} 
\label{eq:DW} 
\Delta _C^W 
= 
i\sum_{\alpha ,\beta = +,-} 
\frac{\delta }{\delta a_{\alpha }} 
G^W_{\alpha \beta} 
\frac{\delta }{\delta a_{\beta }%
^{\dag}%
} 
\, . 
\end{eqnarray}%
The question then is if $\tilde\Delta _C^W$ in (\ref{eq:WickTFDWI}) 
may be replaced 
by $\Delta _C^W$. The same arguments as above show that 
the ``holes'' in 
\mbox{$%
G^W_{-+}(t)%
$} 
and 
\mbox{$%
G^W_{+-}(t)%
$} 
are of no concern. 
This is {\em not\/} the case for 
\mbox{$%
G^W_{++}(t)%
$} 
and 
\mbox{$%
G^W_{--}(t)%
$}%
. 
Indeed, recall that the $T_+$ and $T_-$-ordering are 
both specified for equal times as normal ordering. 
$\tilde\Delta _C^W$ takes care of such same-time operator 
groups reordering them symmetrically, whereas $\Delta _C^W$ 
misses them. 
That is, $\Delta _C^W$ enforces the no-contractions-between-%
the-same-time-operators caveat of Wick's theorem, 
whereas this caveat {\em does not apply\/} if a double-time-%
ordered operator product is reordered symmetrically. We should 
thus either 
amend Wick's theorem introducing same-time contractions, or, 
which is more convenient and suitable for our purposes, 
redefine the time orderings. 
We then obtain 
\begin{eqnarray} 
\nonumber 
&&T_-^W 
\hat P_- 
\ 
T_+^W 
\hat P_+ 
= 
\text{W} 
\left \{ 
\left [ 
\text{e}^{%
\Delta^W_C 
} 
\left ( 
\hat P_-|_{ 
\hat a(t)\rightarrow a_-(t), 
\hat a%
^{\dag}%
(t)\rightarrow a_-%
^{\dag}%
(t) 
} 
\right.\right.\right.\\&&\hspace{5ex}\times\left.\left.\left. 
\hat P_+|_{ 
\hat a(t)\rightarrow a_+(t), 
\hat a%
^{\dag}%
(t)\rightarrow a_+%
^{\dag}%
(t) 
} 
\right ) 
\right ]|_{ 
a_-(t),a_+(t)\rightarrow \hat a(t), 
a_-%
^{\dag}%
(t),a_+%
^{\dag}%
(t)\rightarrow \hat a%
^{\dag}%
(t) 
} 
\right \} 
, 
\label{eq:%
WickTFDW%
} 
\end{eqnarray}%
were $T_-^W$ and $T_+^W$ differ from $T_-$ and $T_+$ in 
that same-time operators are ordered symmetrically rather than 
normally. 
\subsection{Closed perturbative relations for quantum 
field averages} 
\subsubsection{The method of normal ordering} 
For practical purposes, the quantities of interest are 
operator averages rather than the operators themselves. 
We therefore consider a characteristic functional 
of averages of double-time-ordered operator products: 
\begin{eqnarray} 
\label{eq:Xi0} 
\Xi 
\left ( 
\zeta _-,\zeta _+,\zeta_- 
^{\dag} 
,\zeta _+%
^{\dag} 
\right ) = 
\left\langle 
T_- \exp 
\left ( 
\zeta _- \hat {\sl a}%
^{\dag} 
+ 
\zeta_- 
^{\dag} 
\hat {\sl a} 
\right ) 
\, T_+ \exp 
\left ( 
\zeta _+ \hat {\sl a}%
^{\dag} 
+ 
\zeta _+%
^{\dag} 
\hat {\sl a} 
\right ) 
\right\rangle 
. 
\end{eqnarray}%
Angle brackets here define an averaging over the 
Heisenberg $\rho $-matrix of the quantum field, 
(or over the field's initial 
state, which is the same thing): 
\begin{eqnarray} 
\left\langle \cdots\right\rangle = 
\text{Tr} 
\hat \rho 
\left ( 
\cdots\ 
\right ). 
\end{eqnarray}%

Although defined in q-number terms, 
in itself functional (\ref{eq:Xi0}) is a c-number object. 
Applying the closed perturbative relation, 
Eq.\ (\protect\ref{eq:DTP})%
, 
and then Wick's theorem, 
Eq.\ (\protect\ref{eq:%
WickTFD%
})%
, allows one to 
express it as an average of a normally ordered 
operator expression. 
To eliminate q-numbers completely, we shall characterise 
the initial state of the field by the 
corresponding P-distribution, 
\begin{eqnarray} 
P(\alpha ) &=& 
\frac{1}{\pi^2} \int d^2\eta \, 
\left\langle 
\text{e}^{ 
\eta 
\left ( 
\hat a%
^{\dag} 
- \alpha ^* 
\right ) 
- \eta ^* 
\left ( 
\hat a - \alpha 
\right ) - |\eta |^2/2 
}%
\right\rangle 
, 
\\ 
\hat\rho &=& \int d^2 \alpha P(\alpha ) 
\left|%
\alpha 
\right\rangle 
\left\langle 
\alpha 
\right| 
, 
\end{eqnarray}%
where $%
\left|%
\alpha 
\right\rangle%
$ is a coherent state with the 
amplitude $\alpha $: 
\begin{eqnarray} 
\hat a 
\left|%
\alpha 
\right\rangle 
= \alpha 
\left|%
\alpha 
\right\rangle%
. 
\end{eqnarray}%
For any normally ordered operator expression, 
$:\hat X:$, we then have 
\begin{eqnarray} 
\label{eq:AvN} 
\left\langle 
:\hat X:%
\right\rangle 
= \int d^2 \alpha P(\alpha ) 
\left\langle 
\alpha 
\left|%
:\hat X:%
\right|%
\alpha 
\right\rangle 
= 
\int d^2 \alpha P(\alpha )\hat X | 
_{\hat a(t)\to\alpha(t),\hat a%
^{\dag}%
(t)\to\alpha^*(t)} , 
\end{eqnarray}%
where $\alpha(t)= \alpha%
\text{e}^{%
-i\omega t%
}%
$ is the coherent amplitude of 
the interaction-picture field operator, 
$\hat a(t)%
\left|%
\alpha 
\right\rangle%
=\alpha (t) 
\left|%
\alpha 
\right\rangle%
$. 
Combining 
Eqs.\ (\protect\ref{eq:DTP})%
, (\ref{eq:WickTFD}) and (\ref{eq:AvN}) 
then yields: 
\begin{eqnarray} 
\Xi 
\left ( 
\zeta _-,\zeta _+,\zeta_- 
^{\dag} 
,\zeta _+%
^{\dag} 
\right ) &=& 
\int d^2 \alpha P(\alpha )\, 
\text{e}^{%
\Delta _C%
} 
\text{e}^{ 
\zeta _- a_-%
^{\dag} 
+ \zeta _-%
^{\dag} 
a_- 
+ 
\zeta _+ a_+%
^{\dag} 
+ \zeta _+%
^{\dag} 
a_+ 
} 
\nonumber \\ &&\times\,%
\text{e}^{ 
\frac{i\kappa }{4} 
\left ( 
a_-%
^{\dag 2}%
a_-^2 
- 
a_+%
^{\dag 2}%
a_+^2 
\right ) 
} 
|_{a_-(t)=a_+(t)=\alpha(t) ,a_-%
^{\dag}%
(t)=a_+%
^{\dag}%
(t)=\alpha^*(t) } 
\,. 
\label{eq:Xi1} 
\end{eqnarray}%
\subsubsection{The method of Wigner ordering} 
In order to derive an analog of 
Eq.\ (\protect\ref{eq:Xi1}) 
for symmetric ordering, 
consider, to start with, relation (\ref{eq:Ev}) for the evolution 
operator. 
By derivation \cite{QFT}, it is a sum 
of $T$-ordered terms, found by expanding the exponent in a power 
series: 
\begin{eqnarray} 
\nonumber 
{\cal U}(t,t_0) &=& T\exp 
\left [ 
-i \int _{t_0}^t dt' {\cal H}_{\text{int}} (t') 
\right ] 
\\ &=& 
\openone - i \int _{t_0}^t dt' {\cal H}_{\text{int}} (t') + 
\frac{(-i)^2}{2} \int _{t_0}^t dt' dt'' 
T 
{\cal H}_{\text{int}} (t'){\cal H}_{\text{int}} (t'') 
+ \cdots \, . 
\end{eqnarray}%
The $T$-ordering here applies, 
strictly speaking, to the 
Hamiltonian operators regarded as entities rather than to the 
field operators. 
Redefining the time ordering for the field operators via 
Eq.\ (\protect\ref{eq:Tp}) 
changes the integrands, because, e.g., 
\begin{eqnarray} 
T 
{\cal H}_{\text{int}} (t){\cal H}_{\text{int}} (t) 
= 
\left [ 
{\cal H}_{\text{int}} (t) 
\right ]^2 
\neq 
T_+ 
{\cal H}_{\text{int}} (t){\cal H}_{\text{int}} (t) 
= \, 
:\left [ 
{\cal H}_{\text{int}} (t) 
\right ]^2:\,. 
\end{eqnarray}%
This change, however, (i) is finite and (ii) affects only the submanifold 
of measure zero 
of the integration manifold. 
Hence it has no effect on the 
result of integration (this and further arguments below may become 
flawed for 
continuous-space problems where expressions like $\left [ 
{\cal H}_{\text{int}} (t) 
\right ]^2 
- 
:\left [ 
{\cal H}_{\text{int}} (t) 
\right ]^2:$ often contain infinities). 

The key property of the $T_+$-ordering 
is that it does not affect the 
Hamiltonian operators themselves, 
\begin{eqnarray} 
T_+ 
\left \{ 
{\cal H}_{\text{int}} (t) 
\right \} = 
{\cal H}_{\text{int}} (t), 
\end{eqnarray}%
which in turn is due to the normal form of 
${\cal H}_{\text{int}} (t)$. 
It is then immediately clear that (\ref{eq:Ev}) may be rewritten 
in terms of $T_+^W$ by using a symmetric form of the interaction: 
\begin{eqnarray} 
\label{eq:HW} 
\frac{\kappa }{4}\,{\hat{a}^{\dagger 2}}(t) {\hat{a}}^{2}(t) 
= 
\frac{\kappa }{4}\,%
\text{W} 
\left \{ 
{\hat{a}^{\dagger 2}} (t){\hat{a}}^{2}(t) 
- 2 {\hat{a}^{\dagger}}(t) {\hat{a}}(t) + \frac{1}{2} 
\right \} \equiv {\cal H}^W_{\text{int}} (t) 
, 
\end{eqnarray}%
so that 
\begin{eqnarray} 
T_+^W 
\left \{ 
{\cal H}^W_{\text{int}} (t) 
\right \} = 
{\cal H}^W_{\text{int}} (t), 
\end{eqnarray}%
When deriving (\ref{eq:HW}) we have used the property that 
\begin{eqnarray} 
\text{W} 
\left \{ 
{\hat{a}^{\dagger 2}} {\hat{a}}^{2} 
\right \} 
&=& 
\frac{1}{6}\left ( 
{\hat{a}^{\dagger 2}} {\hat{a}}^{2} + \hat{a}^{\dagger} \hat{a} 
\hat{a}^{\dagger} \hat{a} + 
\hat{a}^{\dagger}{\hat{a}}^{2}\hat{a}^{\dagger} + \hat{a} 
{\hat{a}^{\dagger 2}} \hat{a} + \hat{a}\hat{a}^{\dagger} 
\hat{a}\hat{a}^{\dagger} +{\hat{a}}^{2} {\hat{a}^{\dagger 2}} 
\right ), 
\\ 
\text{W} 
\left \{ 
{\hat{a}^{\dagger}} {\hat{a}} 
\right \}&=&\frac{1}{2}\left ( 
{\hat{a}^{\dagger}} {\hat{a}} + \hat{a} 
\hat{a}^{\dagger}\right ) . 
\end{eqnarray} 
For the evolution operator we then find, 
\begin{eqnarray} 
\label{eq:EvW} 
{\cal U}(t,t_0) &=& T_+^W \exp 
\left [ 
-i\int_{t_0}^t dt' {\cal H}^W_{\text{int}} (t') 
\right ] 
\\ 
&=& 
T_+^W 
\exp 
\left \{ 
- \frac{i \kappa }{4} 
\int_{t_0}^{t}dt' 
\left [ 
\hat a%
^{\dag 2}%
(t')\hat a^{2}(t') - 2 
\hat a%
^{\dag}%
(t')\hat a(t')+ \frac{1}{2} 
\right ] 
\right \} 
\label{eq:%
EvWFull%
} 
. 
\end{eqnarray}%
Relation (\ref{eq:DTP}) also remains valid after 
replacing $T_+\to T_+^W$ and ${\cal H}_{\text{int}} (t) 
\to{\cal H}^W_{\text{int}} (t)$. 
For any symmetrically ordered operator expression, 
$%
\text{W} 
\left \{ 
\hat X 
\right \} 
$, we have 
\begin{eqnarray} 
\label{eq:AvW} 
\left\langle 
\text{W} 
\left \{ 
\hat X 
\right \}%
\right\rangle 
= 
\int d^2 \alpha W(\alpha )\hat X | 
_{\hat a(t)\to\alpha(t),\hat a%
^{\dag}%
(t)\to\alpha^*(t)} , 
\end{eqnarray}%
where $W(\alpha )$ is the Wigner distribution 
characterising the initial state, 
\begin{eqnarray} 
W(\alpha ) = 
\frac{1}{\pi^2} \int d^2\eta \, 
\left\langle 
\text{e}^{ 
\eta 
\left ( 
\hat a%
^{\dag} 
- \alpha ^* 
\right ) 
- \eta ^* 
\left ( 
\hat a - \alpha 
\right ) 
}%
\right\rangle 
. 
\end{eqnarray}%
Finally, we arrive at: 
\begin{eqnarray} 
\Xi ^W 
\left ( 
\zeta _-,\zeta _+,\zeta_- 
^{\dag} 
,\zeta _+%
^{\dag} 
\right ) &=& 
\int d^2 \alpha W(\alpha )\, 
\text{e}^{%
\Delta _C^W%
} 
\text{e}^{ 
\zeta _- a_-%
^{\dag} 
+ \zeta _-%
^{\dag} 
a_- 
+ 
\zeta _+ a_+%
^{\dag} 
+ \zeta _+%
^{\dag} 
a_+ 
} 
\nonumber \\ &&\times\,%
\text{e}^{ 
\frac{i\kappa }{4} 
\left ( 
a_-%
^{\dag 2}%
a_-^2 
- 
a_+%
^{\dag 2}%
a_+^2 
+ 2 
a_+%
^{\dag} 
a_+ 
- 2 
a_-%
^{\dag} 
a_- 
\right ) 
} 
|_{a_-(t)=a_+(t)=\alpha(t) ,a_-%
^{\dag}%
(t)=a_+%
^{\dag}%
(t)=\alpha^*(t) } 
\, , 
\label{eq:XiW1} 
\end{eqnarray}%
where 
\begin{eqnarray} 
\label{eq:XiW0} 
\Xi^W 
\left ( 
\zeta _-,\zeta _+,\zeta_- 
^{\dag} 
,\zeta _+%
^{\dag} 
\right ) = 
\left\langle 
T_-^W \exp 
\left ( 
\zeta _- \hat {\sl a}%
^{\dag} 
+ 
\zeta_- 
^{\dag} 
\hat {\sl a} 
\right ) 
\, T_+^W \exp 
\left ( 
\zeta _+ \hat {\sl a}%
^{\dag} 
+ 
\zeta _+%
^{\dag} 
\hat {\sl a} 
\right ) 
\right\rangle 
. 
\end{eqnarray}%

Note that averages generated by functionals 
(\ref{eq:Xi0}) and (\ref{eq:XiW0}) may indeed differ. 
For instance, with $\hat {\sl n}(t) = \hat {\sl a}%
^{\dag}%
(t)\hat {\sl a}(t)$, 
\begin{eqnarray} 
\left\langle 
T_- \hat {\sl n}(t) 
\, T_+ \hat {\sl n}(t) 
\right\rangle 
&=& 
\left\langle  
\left [ 
\hat {\sl n}(t) 
\right ]^2 
\right\rangle%
, 
\\ 
\left\langle 
T_-^W \hat {\sl n}(t) 
\, T_+^W \hat {\sl n}(t) 
\right\rangle 
&=& 
\left\langle  
\left [ 
\hat {\sl n}(t) + \frac{1}{2} 
\right ]^2 
\right\rangle 
. 
\end{eqnarray}%
To obtain $%
\left\langle  
\left [ 
\hat {\sl n}(t) 
\right ]^2 
\right\rangle%
$ directly from (\ref{eq:XiW0}), we may use, e.g., the property 
\begin{eqnarray} 
\left\langle  
\left [ 
\hat {\sl n}(t) 
\right ]^2 
\right\rangle 
= \lim _{\delta t\searrow 0} 
\left\langle 
T_-^W \hat {\sl a}%
^{\dag}%
(t)\hat {\sl a}(t+\delta t) 
\, T_+^W \hat {\sl a}%
^{\dag}%
(t+\delta t)\hat {\sl a}(t) 
\right\rangle 
. 
\end{eqnarray}%
This limit should be preceded by 
the one associated with causal regularisation. 
\section{Causal variables} 
\label{SecCaus} 
Our next goal is to investigate the 
{\em causal structure\/} of 
Eqs.\ (\protect\ref{eq:Xi1}) 
and (\ref{eq:XiW1}). 
The concept of causality is introduced via the 
retarded Green's function, $G(t)$. 
Following this, we are able to define the {\em input and output\/} of 
a quantum system. 
We then show that, physically, the input and output 
thus introduced correspond to 
a generalisation of Kubo's linear reaction approach 
\cite{Kubo} to 
a full nonlinear quantum-stochastic response problem. 
\subsection{The method of normal ordering} 
Consider in more detail the differential quadratic form in 
(\ref{eq:Xi1}). Making use of relations (\ref{eq:GMatrix}), 
and utilising the notation introduced by 
Eq.\ (\protect\ref{eq:N2S2})%
, we find 
\begin{eqnarray} 
\Delta _C = i 
\left ( 
\frac{\delta }{\delta a_+} 
+ 
\frac{\delta }{\delta a_-} 
\right ) 
G \, 
\frac{\delta }{\delta 
a_+%
^{\dag}%
} 
- i 
\left ( 
\frac{\delta }{\delta 
a_+%
^{\dag}%
} 
+ 
\frac{\delta }{\delta 
a_-%
^{\dag}%
} 
\right ) 
G^* \, 
\frac{\delta }{\delta a_-} 
\end{eqnarray}%
We now change the functional variables, 
$a_{\pm}(t),a_{\pm}%
^{\dag}%
(t)\to a(t), a%
^{\dag}%
(t), \xi(t) , \xi 
^{\dag}%
(t)$, 
in order to obtain 
\begin{eqnarray} 
\Delta _C = 
\frac{\delta }{\delta a}%
G%
\frac{\delta }{\delta 
\xi 
^{\dag}%
} 
+ 
\frac{\delta }{\delta a%
^{\dag}%
}%
G^*%
\frac{\delta }{\delta \xi } 
. 
\end{eqnarray}%
That is, 
\begin{eqnarray} 
\frac{\delta }{\delta 
\xi 
^{\dag}%
(t)%
} 
&=&i 
\frac{\delta }{\delta 
a_+%
^{\dag}%
(t)%
} 
, 
\\ 
\frac{\delta }{\delta 
\xi (t)%
} 
&=&-i%
\frac{\delta }{\delta 
a_-(t)%
} 
, 
\\ 
\frac{\delta }{\delta a(t)} 
&=&%
\frac{\delta }{\delta 
a_+(t)%
} 
+ 
\frac{\delta }{\delta 
a_-(t)%
} 
, 
\\ 
\frac{\delta }{\delta 
a%
^{\dag}%
(t)%
} 
&=&%
\frac{\delta }{\delta 
a_+%
^{\dag}%
(t)%
} 
+ 
\frac{\delta }{\delta 
a_-%
^{\dag}%
(t)%
} 
. 
\end{eqnarray}%
These relations determine the new variables up to 
given functions which we chose to be zero: 
\begin{eqnarray} 
\label{eq:%
CausA1%
} 
a_+(t) &=& a(t) , \\ 
a_-(t) &=& a(t) - i \xi (t), \\ 
a_+%
^{\dag}%
(t) &=& a%
^{\dag}%
(t) + i\xi%
^{\dag} 
(t), \\ 
a_-%
^{\dag}%
(t) &=& a%
^{\dag}%
(t) . 
\label{eq:%
CausA4%
} 
\end{eqnarray}%
Consider now the second exponent in (\ref{eq:Xi1}): 
\begin{eqnarray} 
\zeta _- a_-%
^{\dag} 
+ \zeta _-%
^{\dag} 
a_- 
+ 
\zeta _+ a_+%
^{\dag} 
+ \zeta _+%
^{\dag} 
a_+ 
= 
\left ( 
\zeta _- +\zeta _+ 
\right ) a%
^{\dag} 
+ 
\left ( 
\zeta _-%
^{\dag} 
+\zeta _+%
^{\dag} 
\right ) a 
- i \zeta _-%
^{\dag} 
\xi 
+ i \zeta _+ \xi%
^{\dag} 
. 
\label{eq:SecE} 
\end{eqnarray}%
This clearly suggests another substitution, this time 
in the functional $\Xi$ itself: 
\begin{eqnarray} 
\Xi 
\left ( 
\zeta _-,\zeta _+,\zeta _-%
^{\dag}%
,\zeta _+%
^{\dag} 
\right ) \equiv \Phi 
\left ( 
\zeta ,\zeta 
^{\dag}%
,\sigma ,\sigma 
^{\dag} 
\right ) , 
\end{eqnarray}%
where 
\begin{eqnarray} 
\zeta _- (t) +\zeta _+ (t) &=& \zeta (t) , 
\\ 
\zeta _-%
^{\dag} 
(t) +\zeta _+%
^{\dag} 
(t) &=& \zeta 
^{\dag} 
(t) , 
\\ 
- i \zeta _-%
^{\dag} 
(t) &=& \sigma 
^{\dag} 
(t) , 
\\ 
+ i \zeta _+ (t) &=& \sigma (t) , 
\end{eqnarray}%
and 
\begin{eqnarray} 
\label{eq:%
CausZ1%
} 
\zeta _+ (t) &=& - i \sigma (t) , 
\\ 
\zeta _- (t) &=& \zeta (t) + i \sigma (t) , 
\\ 
\zeta _+%
^{\dag} 
(t) &=& \zeta 
^{\dag} 
(t)- i \sigma 
^{\dag} 
(t) , 
\\ 
\zeta _- 
^{\dag}%
(t) &=& i \sigma 
^{\dag} 
(t) . 
\label{eq:%
CausZ4%
} 
\end{eqnarray}%
In {\em causal variables\/} 
thus introduced, 
Eq.\ (\protect\ref{eq:Xi1}) 
becomes: 
\begin{eqnarray} 
\Phi 
\left ( 
\zeta ,\zeta 
^{\dag}%
,\sigma ,\sigma 
^{\dag} 
\right ) &=& 
\int d^2 \alpha P(\alpha ) 
\exp 
\left ( 
\frac{\delta }{\delta a}%
G%
\frac{\delta }{\delta 
\xi 
^{\dag}%
} 
+ 
\frac{\delta }{\delta a%
^{\dag}%
}%
G^*%
\frac{\delta }{\delta \xi } 
\right ) 
\exp 
\left ( 
\zeta a%
^{\dag} 
+ \zeta 
^{\dag} 
a 
+ \xi 
\sigma 
^{\dag} 
+ \xi 
^{\dag} 
\sigma 
\right ) 
\nonumber \\ 
&&\times 
\exp S_{\text{int}} \left ( 
\xi ,\xi 
^{\dag}%
,a ,a 
^{\dag} 
\right ) 
|_{a(t)=\alpha(t) ,a%
^{\dag}%
(t) = \alpha ^*(t), \xi(t) =\xi 
^{\dag}%
(t) =0} 
\, , 
\label{eq:Xi2} 
\end{eqnarray}%
where 
\begin{eqnarray} 
\label{eq:Sint} 
S_{\text{int}} \left ( 
\xi ,\xi 
^{\dag}%
,a ,a 
^{\dag} 
\right ) = 
\frac{\kappa }{2} 
\left ( 
\xi a%
^{\dag 2} 
a + \xi 
^{\dag} 
a^{2} a%
^{\dag} 
\right ) + 
\frac{i\kappa }{4} 
\left ( 
\xi%
^{\dag 2} 
a^{2} 
- 
\xi^2 a%
^{\dag 2} 
\right ) 
. 
\end{eqnarray}%

It should be stressed that 
equation (\ref{eq:Xi2}) is universal, 
while all details of the problem enter 
through $S_{\text{int}} $. 
In general $S_{\text{int}} $ is found as 
\begin{eqnarray} 
S_{\text{int}} \left ( 
\xi ,\xi 
^{\dag}%
,a ,a 
^{\dag} 
\right ) = i \int dt 
\left [ 
h 
\left ( 
a%
^{\dag}%
(t) ,a(t) - i \xi (t) 
\right ) 
- 
h 
\left ( 
a%
^{\dag}%
(t) + i \xi%
^{\dag} 
(t),a(t) 
\right ) 
\right ] 
, 
\end{eqnarray}%
where $h$ is the normal representation of the 
interaction Hamiltonian, 
\begin{eqnarray} 
{\cal H}_{\text{int}} = \ : h 
\left ( 
\hat a%
^{\dag}%
,\hat a 
\right ): 
. 
\end{eqnarray}%
Generalisation of (\ref{eq:Xi2}) to multimode problems 
(see Sec.\ \ref{SecCook}) 
is also straightforward. 
\subsection{The method of Wigner ordering} 
Causal variables for the case of symmetric ordering 
are defined in 
a similar way. 
We require that 
\begin{eqnarray} 
\Delta _C^W = 
\frac{\delta }{\delta a}%
G%
\frac{\delta }{\delta 
\xi 
^{\dag}%
} 
+ 
\frac{\delta }{\delta a%
^{\dag}%
}%
G^*%
\frac{\delta }{\delta \xi } 
, 
\end{eqnarray}%
and then proceed as above to yield two sets of substitutions: 
\begin{eqnarray} 
\label{eq:%
CausAW1%
} 
a_+(t) &=& a(t) + \frac{i}{2}\xi (t) , \\ 
a_-(t) &=& a(t) - \frac{i}{2}\xi (t) , \\ 
a_+%
^{\dag}%
(t) &=& a%
^{\dag}%
(t) + \frac{i}{2}\xi%
^{\dag} 
(t) , \\ 
a_-%
^{\dag}%
(t) &=& a%
^{\dag}%
(t) - \frac{i}{2}\xi%
^{\dag} 
(t), 
\label{eq:%
CausAW4%
} 
\end{eqnarray}%
and 
\begin{eqnarray} 
\label{eq:%
CausZW1%
} 
\zeta _+ (t) &=& \frac{1}{2}\zeta (t) - i \sigma (t) , 
\\ 
\zeta _- (t) &=& \frac{1}{2}\zeta (t) + i \sigma (t) , 
\\ 
\zeta _+%
^{\dag} 
(t) &=& \frac{1}{2} \zeta 
^{\dag} 
(t)- i \sigma 
^{\dag} 
(t) , 
\\ 
\zeta _- 
^{\dag}%
(t) &=& \frac{1}{2} \zeta 
^{\dag} 
(t) + i \sigma 
^{\dag} 
(t) . 
\label{eq:%
CausZW4%
} 
\end{eqnarray}%
Further, on rewriting functional (\ref{eq:XiW0}) in the variables 
(\ref{eq:CausZW1})--(\ref{eq:CausZW4}), 
\begin{eqnarray} 
\Xi^W 
\left ( 
\zeta _-,\zeta _+,\zeta _-%
^{\dag}%
,\zeta _+%
^{\dag} 
\right ) \equiv \Phi^W 
\left ( 
\zeta ,\zeta 
^{\dag}%
,\sigma ,\sigma 
^{\dag} 
\right ) , 
\end{eqnarray}%
we find that 
\begin{eqnarray} 
\Phi ^W 
\left ( 
\zeta ,\zeta 
^{\dag}%
,\sigma ,\sigma 
^{\dag} 
\right ) &=& 
\int d^2\alpha W(\alpha )\exp 
\left ( 
\frac{\delta }{\delta a}%
G%
\frac{\delta }{\delta 
\xi 
^{\dag}%
} 
+ 
\frac{\delta }{\delta a%
^{\dag}%
}%
G^*%
\frac{\delta }{\delta \xi } 
\right ) 
\exp 
\left ( 
\zeta a%
^{\dag} 
+ \zeta 
^{\dag} 
a 
+ \xi 
\sigma 
^{\dag} 
+ \xi 
^{\dag} 
\sigma 
\right ) 
\nonumber \\ 
&&\times 
\exp S^W_{\text{int}} \left ( 
\xi ,\xi 
^{\dag}%
,a ,a 
^{\dag} 
\right ) 
|_{a(t)=\alpha(t) ,a%
^{\dag}%
(t) = \alpha ^*(t), \xi(t) =\xi 
^{\dag}%
(t) =0} 
. 
\label{eq:XiW2} 
\end{eqnarray}%
Here, 
\begin{eqnarray} 
S^W_{\text{int}} \left ( 
\xi ,\xi 
^{\dag}%
,a ,a 
^{\dag} 
\right ) &=& i \int dt 
\left [ 
h^W 
\left ( 
a%
^{\dag}%
(t)- \frac{i}{2}\xi%
^{\dag}%
(t) ,a(t) - \frac{i}{2}\xi (t) 
\right ) 
\right . 
\nonumber \\ && \hspace{10ex} 
\left . 
- \, 
h^W 
\left ( 
a%
^{\dag}%
(t)+ \frac{i}{2}\xi%
^{\dag}%
(t) ,a(t) + \frac{i}{2}\xi (t) 
\right ) 
\right ] 
, 
\end{eqnarray}%
where $h^W$ is the symmetric representation of the 
interaction Hamiltonian, 
\begin{eqnarray} 
{\cal H}_{\text{int}} = 
\text{W} 
\left \{  
h^W 
\left ( 
\hat a%
^{\dag}%
,\hat a 
\right ) 
\right \} 
. 
\end{eqnarray}%
For the Kerr nonlinearity, 
\begin{eqnarray} 
\label{eq:SW} 
S^W_{\text{int}} \left ( 
\xi ,\xi 
^{\dag}%
,a ,a 
^{\dag} 
\right ) = 
\frac{\kappa }{2} 
\left [ 
\left ( 
\xi a%
^{\dag} 
+ \xi 
^{\dag} 
a 
\right ) 
\left ( 
a%
^{\dag} 
a -2 
\right ) 
\right ] - 
\frac{\kappa }{8} 
\left ( 
\xi^2 \xi%
^{\dag} 
a%
^{\dag} 
+ \xi%
^{\dag 2} 
\xi a 
\right ) 
. 
\end{eqnarray}%

It should not be overlooked that the definition 
of causal variables is ordering-specific. 
In fact, the notation $\zeta ,\zeta 
^{\dag}%
,\sigma ,\sigma 
^{\dag}%
$ 
is used for two different sets of causal variables: 
(\ref{eq:CausZ1})--(\ref{eq:CausZ4}) 
in the case of normal ordering, and 
(\ref{eq:CausZW1})--(\ref{eq:CausZW4}) 
in the case of symmetric ordering. 
Since it is always clear by the context which case is 
being discussed, no confusion should occur. 
\subsection{Quantum nonlinear-reaction problem} 
To gain more insight into the causal variables, 
we now introduce a variable pump term into 
the interaction Hamiltonian, which then reads: 
\begin{eqnarray} 
\tilde H_{\text{int}} (t) = \frac{\kappa }{4} 
\hat a%
^{\dag 2} 
(t)\hat a^2(t) 
+ s(t)\hat a%
^{\dag} 
(t)+ s^*(t)\hat a (t) . 
\end{eqnarray}%
The {\em external source\/}, $s(t)$, is a 
given c-number function. 
We mark by tilde all quantitities defined in the presence 
of the source. 
With the source, 
\begin{eqnarray} 
\tilde S_{\text{int}} = S_{\text{int}} + \xi s^* + \xi 
^{\dag} 
s . 
\end{eqnarray}%
Moving $\xi s^* + \xi 
^{\dag} 
s$ to the second exponent in 
(\ref{eq:Xi2}) results in an identity, 
\begin{eqnarray} 
\tilde \Phi 
\left ( 
\zeta ,\zeta 
^{\dag}%
,\sigma ,\sigma 
^{\dag} 
\right ) = \Phi 
\left ( 
\zeta ,\zeta 
^{\dag}%
,\sigma + s ,\sigma 
^{\dag} 
+ s^* 
\right ) , 
\end{eqnarray}%
which also holds for the $\tilde\Phi^W/\Phi^W$ pair. 

This way, the physical information contained in the dependence of 
both $\Phi\left ( 
\zeta ,\zeta 
^{\dag}%
,\sigma ,\sigma 
^{\dag} 
\right )$ and $\Phi^W\left ( 
\zeta ,\zeta 
^{\dag}%
,\sigma ,\sigma 
^{\dag} 
\right )$ on $\sigma $ and $\sigma 
^{\dag}%
$ 
is the system's reaction to an external perturbations, while 
the variables $\sigma ,\sigma 
^{\dag}%
$ define an input of the system. 
The fact that this input is also determined dynamically makes 
it independent of ordering. 
The variables $\zeta ,\zeta 
^{\dag}%
$ define an output of the system. 
Unlike the input, the output turns out to be ordering-specific. Consider 
firstly the case of normal ordering; 
assuming the source to be arbitrary, 
the full physical information 
about the system is obtainable from 
\begin{eqnarray} 
\label{eq:TPhi} 
\tilde \Phi 
\left ( 
\zeta ,\zeta 
^{\dag}%
,0,0 
\right ) = 
\Phi 
\left ( 
\zeta ,\zeta 
^{\dag}%
, s , s^* 
\right ) 
. 
\end{eqnarray}%
In turn, making use of (\ref{eq:Xi0}) we get: 
\begin{eqnarray} 
\label{eq:TN} 
\tilde \Phi 
\left ( 
\zeta ,\zeta 
^{\dag}%
,0,0 
\right ) = 
\left\langle 
T_- \exp 
\left ( 
\zeta 
^{\dag} 
\hat{\tilde{\sl a}} 
\right ) 
\, T_+ \exp 
\left ( 
\zeta \hat{\tilde{\sl a}} 
^{\dag} 
\right ) 
\right\rangle 
. 
\end{eqnarray}%
This is nothing but a characteristic functional of 
Glauber's renowned time-normal averages of the Heisenberg 
field operator in the presence of the source. 
The source terms in the Hamiltonian are also quite recognisable; 
they appear in Kubo's linear reaction theory \cite{Kubo}. 
Introducing causal variables is thus equivalent to a nonlinear-%
reaction reformulation of a quantum system. 

Unlike the linear reaction theory, the nonlinear reaction theory 
depends explicitly on which quantities are to be 
``measured''. 
Causal variables for the normal ordering, 
Eqs.\ (\protect\ref{eq:%
CausZ1%
})%
--%
(\ref{eq:CausZ4}), correspond to ``measuring'' time-normal 
averages of the field operators. 
Causal variables for the symmetric ordering, 
Eqs.\ (\protect\ref{eq:%
CausZW1%
})%
--%
(\ref{eq:CausZW4}), introduce another set of ``measured'' 
quantities. 
For example, 
relation (\ref{eq:TPhi}) holds equally 
for $\tilde\Phi^W$ and $\Phi ^W$ 
and in place of (\ref{eq:TN}), one finds 
\begin{eqnarray} 
\label{eq:TW} 
\tilde \Phi^W 
\! \left ( 
\zeta ,\zeta 
^{\dag}%
,0,0 
\right ) = 
\left\langle 
T_- ^W\exp 
\left ( 
\frac{1}{2} 
\zeta 
^{\dag} 
\hat{\tilde{\sl a}} 
+ \frac{1}{2} 
\zeta \hat{\tilde{\sl a}} 
^{\dag} 
\right ) 
\, T_+ ^W\exp 
\left ( 
\frac{1}{2} 
\zeta 
^{\dag} 
\hat{\tilde{\sl a}} 
+ \frac{1}{2} 
\zeta \hat{\tilde{\sl a}} 
^{\dag} 
\right ) 
\right\rangle 
. 
\end{eqnarray}%
Disregarding the time orderings turns (\ref{eq:TW}) into 
a characteristic functional of symmetrically ordered operator 
averages. 
We shall therefore term the operator ordering introduced 
by (\ref{eq:TW}) a time-Wigner ($T_W$) ordering. 
That is, by definition, 
\begin{eqnarray} 
\label{eq:TWA} 
\tilde \Phi^W 
\! \left ( 
\zeta ,\zeta 
^{\dag}%
,0,0 
\right ) &=& 
\left\langle  
T_W\exp 
\left ( 
\zeta 
^{\dag} 
\hat{\tilde{\sl a}} 
+ 
\zeta \hat{\tilde{\sl a}} 
^{\dag} 
\right ) 
\right\rangle%
, 
\end{eqnarray}%
where 
\begin{eqnarray} 
T_W\exp 
\left ( 
\zeta 
^{\dag} 
\hat{{\sl a}} 
+ 
\zeta \hat{{\sl a}} 
^{\dag} 
\right ) 
&=& 
T_- ^W\exp 
\left ( 
\frac{1}{2} 
\zeta 
^{\dag} 
\hat{{\sl a}} 
+ \frac{1}{2} 
\zeta \hat{{\sl a}} 
^{\dag} 
\right ) 
\, T_+ ^W\exp 
\left ( 
\frac{1}{2} 
\zeta 
^{\dag} 
\hat{{\sl a}} 
+ \frac{1}{2} 
\zeta \hat{{\sl a}} 
^{\dag} 
\right ) 
\label{eq:TWO} 
. 
\end{eqnarray}%
We have dropped the tildes here 
to emphasise that, in itself, the 
definition of the time-Wigner ordering does not 
depend on the presence or absence of the sources 
(nor on any other details of the dynamics). 

Under time-normal ordering, 
the ``most recent'' creation (annihilation) operator 
becomes the leftmost (rightmost) in the product. 
The $T_W$-ordering acts in a similar manner, only 
symmetrised in respect of the creation and annihilation 
operators: 
\begin{eqnarray} 
T_W \left\{ 
\hat {\sl x}(t)\hat {\sl y}(t')\cdots\hat {\sl z}(t'') 
\right\} &=& 
\frac{1}{2}\left[ 
T_W \left\{ 
\hat {\sl y}(t')\cdots\hat {\sl z}(t'') 
\right\}\hat {\sl x}(t) 
\right. \nonumber \\ 
&& + \, \left. 
\hat {\sl x}(t) T_W \left\{ 
\hat {\sl y}(t')\cdots\hat {\sl z}(t'') 
\right\} 
\right] 
. 
\label{eq:TWDef} 
\end{eqnarray}%
Here, $\hat {\sl x},\hat {\sl y},\cdots,\hat {\sl z}$ 
stand for field operators 
(i.e., $\hat{\sl a}$ or $\hat{\sl a}%
^{\dag}%
$), 
and $t$ should exceed 
all other time arguments in the product, 
$t>t',\cdots,t''$. 
Equation (\ref{eq:TWDef}) is a recurrence relation, 
defining $T_W$ for the case of time arguments which are all 
different. 
For products of two 
operators, time-Wigner ordering is merely symmetric 
ordering, 
\begin{eqnarray} 
T_W \hat {\sl x}(t) \hat {\sl y}(t') = 
\frac{1}{2} 
\left[ 
\hat {\sl x}(t) \hat {\sl y}(t') 
+ 
\hat {\sl y}(t')\hat {\sl x}(t) 
\right] 
= 
\frac{1}{2} 
\left[ 
\hat {\sl x}(t), \hat {\sl y}(t') 
\right]_+ , 
\end{eqnarray}%
where $[\cdots]_+$ denotes an anticommutator. 
For higher-order products, time ordering becomes 
essential, so that, for three operators, assuming that $t>t',t''$, 
\begin{eqnarray} 
\label{eq:3O} 
T_W \hat {\sl x}(t) \hat {\sl y}(t') \hat {\sl z}(t'') = 
\frac{1}{4} [ 
\hat {\sl x}(t),[ 
\hat {\sl y}(t'),\hat {\sl z}(t'') 
]_+ 
]_+ . 
\end{eqnarray}%
It is easy to see that the following rule holds 
for an arbitrary $T_W$-ordered product: 
write all possible permutations of the operators in a product, 
then retain only those where time arguments are arranged in 
a C-contour sequence (i.e., times increase then decrease). 

Consider now the behaviour of time-Wigner averages for 
equal time arguments. 
Assuming $t'>t$, from (\ref{eq:3O}) we find, 
\begin{eqnarray} 
T_W \hat {\sl x}(t) \hat {\sl y}(t') \hat {\sl z}(t'+0) 
- 
T_W \hat {\sl x}(t) \hat {\sl y}(t') \hat {\sl z}(t'-0) 
= 
\frac{1}{4} [ 
\hat {\sl x}(t),[ 
\hat {\sl y}(t'),\hat {\sl z}(t') 
] 
]_+ 
, 
\end{eqnarray}%
where $[\cdots]$ is a commutator, and 
$T_W\hat {\sl x}(t) \hat {\sl y}(t') \hat {\sl z}(t'\pm 0) \equiv 
\lim_{\delta t\searrow 0} 
T_W\hat {\sl x}(t) \hat {\sl y}(t') \hat {\sl z}(t'\pm \delta t)$ 
(i.e.\ the limit applies to the $T_W$-ordered product as a whole). 
This means that, unlike time-normal products, time-Wigner 
products are not continuous at coinciding times, 
nor do the limits of $T_W$-ordered products at equal times 
coincide with symmetric products. 
For instance, 
\begin{eqnarray} 
T_W 
\hat {\sl a}%
^{\dag} 
(t+0)\hat {\sl a}^2 (t) &=& 
\frac{1}{2} \left[ 
\hat {\sl a}%
^{\dag} 
(t)\hat {\sl a}^2 (t) + 
\hat {\sl a}^2 (t)\hat {\sl a}%
^{\dag} 
(t) 
\right] 
\\ 
\neq%
\text{W} 
\hat {\sl a}%
^{\dag} 
(t)\hat {\sl a}^2 (t) &=& 
\frac{1}{3} \left[ 
\hat {\sl a}%
^{\dag} 
(t)\hat {\sl a}^2 (t) + 
\hat {\sl a} (t)\hat {\sl a}%
^{\dag} 
(t)\hat {\sl a} (t) + 
\hat {\sl a}^2 (t)\hat {\sl a}%
^{\dag} 
(t) 
\right]. 
\end{eqnarray}%
Symmetric averages may only be recovered as combinations of such 
limits from different directions, e.g., 
\begin{eqnarray} 
\text{W} 
\hat {\sl a}%
^{\dag} 
(t)\hat {\sl a}^2 (t) = 
\frac{1}{3} 
T_W 
\hat {\sl a}%
^{\dag} 
(t+0)\hat {\sl a}^2 (t) 
+ 
\frac{2}{3} 
T_W 
\hat {\sl a}%
^{\dag} 
(t)\hat {\sl a} (t)\hat {\sl a} (t+0) 
. 
\end{eqnarray}%
\section{Calculating quantum averages as classical stochastic 
averages} 
\label{SecStoch} 
In this section, we show that there exist classical 
stochastic problems, such that 
Eqs.\ (\protect\ref{eq:Xi1}) 
and (\ref{eq:XiW1}) 
may be interpreted as stochastic averages. 
In particular, 
Eq.\ (\protect\ref{eq:Xi1}) 
yields a 
generalisation of the positive-P representation 
(well known in quantum optics) to a quantum nonlinear response 
problem. 
\subsection{Classical stochastic response problem} 
Relations (\ref{eq:Xi2}) and 
(\ref{eq:XiW2}) give formal solutions to 
quantum response problems 
(formulated, respectively, in terms of time-normal 
and time-Wigner averages). 
It is instructive to compare these relations to 
a solution to a classical stochastic response problem. 
To this end, consider a c-number stochastic field $a(t)$, 
which obeys an integral equation, 
\begin{eqnarray} 
\label{eq:IntEq} 
a(t) = \alpha (t) + \int dt' G(t-t') \sigma_{\text{tot}} (t') 
, 
\end{eqnarray}%
where $\alpha (t)$ is the in-field and 
$\sigma_{\text{tot}} (t)$ 
is the field source. 
For the purposes of this 
paragraph, the kernel $G(t)$ is assumed to be 
regular and retarded, $G(t)=0, t\leq 0$, and otherwise arbitrary. 
We assume that the in-field, $\alpha (t)$, is also arbitrary. 
(This allows one to regularise $G(t)$ 
without changing $\alpha (t)$.) 
The full field source $\sigma_{\text{tot}} (t)$ consists 
of two parts, 
\begin{eqnarray} 
\sigma_{\text{tot}} (t) = 
\sigma (t) + 
\sigma' (t) 
. 
\end{eqnarray}%
The {\em external\/} source, $\sigma (t)$, is regarded 
as given, while the {\em random source\/}, $\sigma' (t)$, 
depends on the field. 
That is, the random source 
describes the field's effective self-action (which physically 
originates, e.g., in interaction with a medium). 
As a random quantity, $\sigma' (t)$ is fully 
characterised by a probability distribution, 
conditional on the field at the same time $t$: 
\begin{eqnarray} 
\label{eq:%
PiCond%
} 
\Pi 
\left ( 
\left . 
\sigma '(t) 
\right | 
a(t) 
\right ) 
. 
\end{eqnarray}%
Resolving formally the self-action problem 
results in a probability distribution over the 
random source, $\sigma' (t)$, 
conditional on the in-field, $\alpha (t)$, 
and the external 
source, $\sigma (t)$. 
This is found by substituting 
Eq.\ (\protect\ref{eq:IntEq}) 
for $a(t)$, 
\begin{eqnarray} 
\label{eq:%
PiUncond%
} 
\Pi 
\left ( 
\left . 
\sigma '(t) 
\right | 
\left [ 
\alpha + G(\sigma +\sigma ') 
\right ] 
(t) 
\right ) 
. 
\end{eqnarray}%
Importantly, this expression does not 
contain a vicious cycle because of the 
assumed regular-and-retarded nature of $G(t)$: 
$\sigma '(t)$ depends on $\sigma '(t')$ only for $t'<t$. 

Consider now statistical properties of the field. 
With the self-action resolved, 
these are also conditional on the in-field, $\alpha (t)$, 
and the external 
source, $\sigma (t)$. 
For the characteristic functional of multi-time 
stochastic field averages we find, 
(with $\zeta (t)$ being an arbitrary function) 
\begin{eqnarray} 
\Sigma 
\left ( \left . 
\zeta \right | \alpha ,\sigma 
\right ) &=& 
\overline{\hspace{0.1ex} 
\exp 
\left ( 
\int dt \zeta (t) a(t) 
\right ) 
\hspace{0.1ex}} 
= 
\overline{\hspace{0.1ex} 
\text{e}^{ 
\zeta 
\left [ 
\alpha + G(\sigma +\sigma ') 
\right ] 
} 
\hspace{0.1ex}} 
\\ 
&=& 
\int \text{D}^{\infty}\sigma '\, 
\text{e}^{ 
\zeta 
\left [ 
\alpha + G(\sigma +\sigma ') 
\right ] 
} 
\Pi 
\left ( 
\left . 
\sigma ' 
\right | 
\alpha + G[\sigma +\sigma '] 
\right ) 
. 
\end{eqnarray}%
The upper bar here denotes an averaging over the 
statistics of $\sigma '$, which is afterwards explicitly 
rewritten as a trajectorial (functional) integral. 
(Here and in what follows, 
we will make extensive use of a brief notation 
so as to prevent our relations from growing bulky.) 
Then, firstly, we 
pull $G(\sigma +\sigma ')$ 
out of $\Pi 
\left ( 
\left . 
\sigma ' 
\right | 
\alpha + G[\sigma +\sigma '] 
\right ) 
$ by applying a shift operator: 
\begin{eqnarray} 
\Pi 
\left ( 
\left . 
\sigma ' 
\right | 
\alpha + G[\sigma +\sigma '] 
\right ) = 
\text{e}^{ 
\frac{\delta }{\delta \alpha } 
G(\sigma +\sigma ')%
} 
\Pi 
\left ( 
\left . 
\sigma ' 
\right | 
\alpha 
\right ) 
. 
\end{eqnarray}%
Secondly, we pull all factors except $\Pi $ 
out of the functional integral, resulting in: 
(with $\xi (t)$ being another arbitrary function) 
\begin{eqnarray} 
\Sigma 
\left ( \left . 
\zeta \right | \alpha ,\sigma 
\right ) 
&=& 
\int \text{D}^{\infty}\sigma '\, 
\text{e}^{ 
\zeta \alpha + 
\left ( 
\zeta + \frac{\delta }{\delta \alpha } 
\right )G 
\left ( 
\sigma + \sigma ' 
\right ) 
} 
\Pi 
\left ( 
\left . 
\sigma ' 
\right | 
\alpha 
\right ) 
\\ 
&=& 
\text{e}^{ 
\zeta \alpha + 
\left ( 
\zeta + \frac{\delta }{\delta \alpha } 
\right )G 
\left ( 
\sigma +\frac{\delta }{\delta \xi } 
\right ) 
} 
\int \text{D}^{\infty}\sigma '\, 
\text{e}^{%
\xi \sigma '%
} 
\Pi 
\left ( 
\left . 
\sigma ' 
\right | 
\alpha 
\right ) 
|_{\xi =0} 
\\ 
&=& 
\text{e}^{ 
\zeta \alpha + 
\left ( 
\zeta + \frac{\delta }{\delta \alpha } 
\right )G 
\left ( 
\sigma +\frac{\delta }{\delta \xi } 
\right ) 
} 
\text{e}^{%
S 
\left ( \left . 
\zeta \right |\alpha 
\right ) 
} 
|_{\xi =0} 
. 
\end{eqnarray}%
We have introduced a characteristic functional 
of {\em cumulants\/} of the random source conditional on the 
full field, 
\begin{eqnarray} 
S 
\left ( \left . 
\zeta \right |a 
\right )  
= \ln\int \text{D}^{\infty}\sigma '\, 
\text{e}^{%
\xi \sigma '%
} 
\Pi 
\left ( 
\left . 
\sigma ' 
\right | 
a 
\right ) 
. 
\end{eqnarray}%
After further algebra, which is much assisted by 
Eq.\ (\protect\ref{eq:PDiff})%
, we arrive at 
\begin{eqnarray} 
\label{eq:1F} 
\Sigma 
\left ( \left . 
\zeta \right | \alpha ,\sigma 
\right ) = 
\text{e}^{%
\frac{\delta }{\delta a} 
G 
\frac{\delta }{\delta \xi }%
} 
\text{e}^{%
\zeta a + \xi \sigma 
} 
\text{e}^{%
S 
\left ( \left . 
\zeta \right |a 
\right ) 
} 
|_{\xi =0, a = \alpha } 
\, . 
\end{eqnarray}%

Disregarding the averaging over the pseudo-distributions, 
Eqs.\ (\protect\ref{eq:Xi2}) 
and (\ref{eq:XiW2}) look very much like generalisations 
of 
Eq.\ (\protect\ref{eq:1F}) 
to a pair of random fields. 
It is not however immediately obvious that 
$S_{\text{int}} $ and $S^W_{\text{int}} $ 
may be interpreted as characteristic functionals 
of cumulants of some classical noises. 
In fact, 
this turns out to be unconditionally the case for 
$S_{\text{int}} $, 
whereas for $S^W_{\text{int}} $ this 
interpretation holds only with discretisation of the time axis. 
\subsection{Hubbard-Stratonovich transformation: 
introducing noise sources constructively} 
\subsubsection{Generalising the positive-P representation to 
a quantum problem of nonlinear reaction} 
Consider a standardised real delta-correlated Gaussian 
noise, $\chi (t)$, 
\begin{eqnarray} 
\overline{\hspace{0.1ex}%
\chi (t)\chi (t')%
\hspace{0.1ex}} 
= \delta (t-t'). 
\end{eqnarray}%
The following relation (Hubbard-Stratonovich Transformation---HST) 
holds for an arbitrary function, $x(t)$, 
\begin{eqnarray} 
\label{eq:HSTR} 
\overline{\hspace{0.1ex}%
\text{e}^{ 
\chi x 
}%
\hspace{0.1ex}} 
= 
\text{e}^{x^2/2} 
, 
\end{eqnarray}%
so we find, 
\begin{eqnarray} 
\label{eq:%
SWNoise%
} 
\exp S_{\text{int}} 
\left ( 
\xi ,\xi 
^{\dag}%
,a,a%
^{\dag} 
\right ) = 
\overline{\hspace{0.1ex} 
\exp 
\left [ 
\xi 
\left ( 
\frac{\kappa }{2} a%
^{\dag 2} 
a 
+ \sqrt{\frac{-i\kappa }{2}}\, 
\chi 
^{\dag} 
a%
^{\dag} 
\right ) 
+ 
\xi 
^{\dag} 
\left ( 
\frac{\kappa }{2} a^{2} a%
^{\dag} 
+ \sqrt{\frac{i\kappa }{2}}\, 
\chi a 
\right ) 
\right ] 
\hspace{0.1ex}}%
, 
\end{eqnarray}%
where $\chi%
^{\dag} 
(t)$ is another standardised 
real delta-correlated Gaussian 
noise, uncorrelated with $\chi (t)$, 
and the upper bar denotes averaging over the 
statistics of the noises. 
The exponent on the RHS of (\ref{eq:SWNoise}) is linear in 
$\xi(t) $ and $\xi 
^{\dag}%
(t)$. 
This means that for given $\chi (t)$ and $\chi 
^{\dag}%
(t)$, 
the equations for $a(t)$ and $a%
^{\dag}%
(t)$ become regular, 
with the random sources being, respectively, 
\begin{eqnarray} 
\sigma '(t) &=& 
\frac{\kappa }{2} a^{2}(t) a%
^{\dag}%
(t) 
+ \sqrt{\frac{i\kappa }{2}}\, 
\chi(t) a(t) , 
\\ 
\sigma%
^{\dag 
\prime%
}%
(t) &=& 
\frac{\kappa }{2} a%
^{\dag 2}%
(t) a(t) 
+ \sqrt{\frac{-i\kappa }{2}}\, 
\chi 
^{\dag}%
(t) a%
^{\dag}%
(t) 
. 
\end{eqnarray}%
In other words, $a(t)$ and $a%
^{\dag}%
(t)$ obey a pair of coupled 
stochastic integral equations, 
\begin{eqnarray} 
a(t) &=& \alpha (t) + 
\int dt' G(t-t') 
\left [ \sigma (t) + 
\frac{\kappa }{2} a^{2}(t') a%
^{\dag}%
(t') 
+ \sqrt{\frac{i\kappa }{2}}\, 
\chi (t') a(t') 
\right ], 
\\ 
a%
^{\dag}%
(t) &=& \alpha^* (t) + 
\int dt' G^*(t-t') 
\left [ \sigma%
^{\dag} 
(t) + 
\frac{\kappa }{2} a%
^{\dag 2}%
(t') a(t') 
+ \sqrt{\frac{-i\kappa }{2}}\, 
\chi%
^{\dag} 
(t') a%
^{\dag} 
(t') 
\right ] 
. 
\end{eqnarray}%
On removing the causal regularisation of $G(t)$, 
one recovers a pair of It\^o stochastic differential 
equations, 
\begin{eqnarray} 
\label{eq:PP1} 
i \, \frac{da(t)}{dt} &=& 
\sigma (t) + 
\frac{\kappa }{2} a^{2}(t') a%
^{\dag}%
(t') 
+ \sqrt{\frac{i\kappa }{2}}\, 
\chi (t') a(t') 
, 
\\ 
-i \, \frac{da%
^{\dag}%
(t)}{dt}&=& 
\sigma%
^{\dag} 
(t) + 
\frac{\kappa }{2} a%
^{\dag 2}%
(t') a(t') 
+ \sqrt{\frac{-i\kappa }{2}}\, 
\chi%
^{\dag} 
(t') a%
^{\dag} 
(t') 
. 
\label{eq:PP2} 
\end{eqnarray}%
The fact that It\^o calculus should be chosen is due 
to the causal regularisation of $G(t)$, 
which makes sources at time $t$ independent of 
fields at the same time, 
which is the characteristic property of It\^o calculus. 

Without the external sources 
(i.e., with $\sigma =\sigma 
^{\dag} 
=0$), 
equations (\ref{eq:PP1}), (\ref{eq:PP2}) are 
well known in quantum optics under the name of the 
positive-P representation. 
In that form, they allow one to calculate time-normal 
averages of the field operators, 
describing {\em radiation\/} properties of the system. 
With sources included, they become applicable to a much wider 
assemblage of operator averages, covering the full 
quantum-stochastic nonlinear {\em reaction\/} problem. 
\subsubsection{Positive-W representation} 
At a first glance, factorising noise terms 
in $S^W_{\text{int}} $ does not pose a problem either. 
Introducing a standardised 
complex delta-correlated Gaussian 
noise, $\eta (t)$, such that 
\begin{eqnarray} 
\overline{\hspace{0.1ex}%
\eta (t)\eta (t')%
\hspace{0.1ex}}%
=0 
, \ \ \ 
\overline{\hspace{0.1ex}%
\eta (t)\eta ^* (t')%
\hspace{0.1ex}}%
=\delta (t-t') 
, 
\end{eqnarray}%
allows one to factorise arbitrary products: 
\begin{eqnarray} 
\label{eq:HSTC} 
\text{e}^{xy} 
= 
\overline{\hspace{0.1ex} 
\text{e}^{%
x\eta + y\eta ^*%
} 
\hspace{0.1ex}}%
. 
\end{eqnarray}%
We shall write the real and complex HST's, 
Eqs.\ (\protect\ref{eq:HSTR}) 
and (\ref{eq:HSTC}), as: 
\begin{eqnarray} 
\label{eq:%
HSTArr%
} 
\frac{x^2}{2} 
\stackrel{\chi}{\longrightarrow} 
x\chi 
, \ \ \ 
xy 
\stackrel{\eta }{\Longrightarrow} 
x\eta + y \eta ^* 
. 
\end{eqnarray}%
These relations imply the definition of the corresponding noises 
as standardised Gaussian noises 
(respectively, real and complex ones). 
Then, e.g., (cf. 
Eq.\ (\protect\ref{eq:SW})%
) 
\begin{eqnarray} 
\label{eq:Naive} 
- \frac{\kappa }{8} \xi%
^{\dag 2}%
\xi a 
\stackrel{\eta}{\Longrightarrow} 
\eta \frac{\xi%
^{\dag 2}%
}{2} - 
\eta ^* \frac{\kappa }{4} \xi a 
\stackrel{\chi }{\longrightarrow} 
\chi \sqrt{\eta}\, \xi 
^{\dag} 
- 
\eta ^* \frac{\kappa }{4} \xi a 
. 
\end{eqnarray}%
Consider now the random quantity, 
$\nu(t) = \chi(t) \sqrt{\eta(t)}\,$. 
$\chi (t)$ and $\eta (t)$ are independent, and we have, 
\begin{eqnarray} 
\overline{\hspace{0.1ex}%
\nu (t)\nu^* (t')%
\hspace{0.1ex}} 
= 
\overline{\hspace{0.1ex}%
\chi (t)\chi (t')%
\hspace{0.1ex}}%
\hspace{0.5ex} 
\overline{\hspace{0.1ex}%
\sqrt{\eta (t)\eta^*(t')}%
\hspace{0.1ex}} 
= 
\delta (t-t') 
\sqrt{ 
\pi \delta (t-t') 
} = 
\delta (t-t') 
\sqrt{ 
\pi \delta (0) 
}\,. 
\end{eqnarray}%
That is, we have encounted an infinity (divergence). 
The underlying reason is certainly that both 
$\chi (t)$ and $\eta (t)$ are highly singular 
functions and the ``quantity'' $\nu (t)$ is 
simply not defined. 

For all practical purposes (such as computer simulations) 
stochastic differential equations will be replaced by 
finite-difference equations. 
We therefore assume an equidistant 
discretisation of the time axis, 
with $\Delta t$ being the step of the time grid. 
The integration and the $\delta$-function are understood as 
usual, 
\begin{eqnarray} 
\int dt \equiv \Delta t \sum_t, \ \ \ 
\delta (t-t') \equiv \Delta t^{-1} \delta _{tt'}, 
\end{eqnarray}%
where $\delta _{tt'}$ is the Kronecker symbol on the time 
grid (i.e.\ $\delta _{tt'}=1$ if $t=t'$ and 0 otherwise). 
We renormalise the standardised 
Gaussian sources, 
\begin{eqnarray} 
\eta (t) = \Delta t^{-1/2} \bar \eta (t), \ \ \ 
\chi (t) = \Delta t^{-1/2} \bar \chi (t) 
, 
\end{eqnarray}%
making them $\delta _{tt'}$ 
(Kronecker symbol) correlated, not $\delta (t-t')$ 
($\delta $-function) correlated: 
\begin{eqnarray} 
\overline{\hspace{0.1ex}%
\bar\eta (t)\bar\eta ^*(t')%
\hspace{0.1ex}} 
= \delta _{tt'}, 
\ \ \ 
\overline{\hspace{0.1ex}%
\bar\chi (t)\bar\chi (t')%
\hspace{0.1ex}} 
= \delta _{tt'} 
. 
\end{eqnarray}%
The short upper bar always marks the Kronecker-correlated noises 
(and should not be confused with the long one denoting the 
averaging; the latter always extends 
over the whole expression). 
The HSTs then become, 
\begin{eqnarray} 
\frac{x^2}{2} 
\stackrel{%
\bar\chi%
}{\longrightarrow} 
\frac{x\bar\chi}{\sqrt{\Delta t}} 
, \ \ \ 
xy 
\stackrel{%
\bar\eta 
}{\Longrightarrow} 
\frac{x\bar\eta + y \bar\eta ^*}{\sqrt{\Delta t}} 
. 
\end{eqnarray}%

There are at least two ways in which the third order terms 
may be factorised. 
One of them, generalising (\ref{eq:Naive}), 
employs two complex and two real HST's: 
\begin{eqnarray} 
\label{eq:Wise} 
&& 
- \frac{\kappa }{8} \xi%
^{\dag 2}%
\xi a 
\stackrel{%
\bar\eta%
}{\Longrightarrow} 
p \Delta t \bar\eta \frac{\xi%
^{\dag 2}%
}{2} + 
q \bar\eta ^* \xi 
\stackrel{%
\bar\chi 
}{\longrightarrow} 
\bar\chi \sqrt{p\bar\eta}\, \xi 
^{\dag} 
+ 
q\bar\eta ^* \xi 
,\\ 
\label{eq:WiseD} 
&& 
- \frac{\kappa }{8} \xi^{2}\xi 
^{\dag} 
a%
^{\dag} 
\stackrel{%
\bar\eta%
^{\dag}%
}{\Longrightarrow} 
p%
^{\dag} 
\Delta t \bar\eta%
^{\dag} 
\frac{\xi^{2}}{2} + 
q%
^{\dag} 
\bar\eta%
^{\dag *} 
\xi%
^{\dag} 
\stackrel{%
\bar\chi%
^{\dag} 
}{\longrightarrow} 
\bar\chi%
^{\dag} 
\sqrt{p%
^{\dag}%
\bar\eta%
^{\dag}%
}\, \xi + 
q%
^{\dag}%
\bar\eta%
^{\dag *} 
\xi 
^{\dag} 
, 
\end{eqnarray}%
where the functions $p(t),q(t),p%
^{\dag}%
(t),q%
^{\dag}%
(t)$ obey the 
relations 
\begin{eqnarray} 
\label{eq:PQ} 
p(t)q(t) = -\frac{\kappa a(t)}{4\Delta t^2} , 
\ \ \ 
p%
^{\dag}%
(t)q%
^{\dag}%
(t) = -\frac{\kappa a%
^{\dag}%
(t)}{4\Delta t^2} , 
\end{eqnarray}%
and are otherwise arbitrary (they can even be random, given 
they are not correlated with the Gaussian noises, 
$\bar\eta(t) ,\bar\chi(t) ,\bar\eta%
^{\dag}%
(t) ,\bar\chi%
^{\dag}%
(t) $). 
This results in the following coupled finite-difference 
equations for $a(t)$ and $a%
^{\dag}%
(t)$: 
\begin{eqnarray} 
\label{eq:WpA} 
i 
\left [ 
a(t+\Delta t)-a(t) 
\right ] 
&=& 
\left \{ 
\frac{\kappa }{2} 
\left [ 
a%
^{\dag}%
(t) a(t) - 2 
\right ]a(t) + \mu (t) 
\right \} \Delta t 
, 
\\ 
-i 
\left [ 
a%
^{\dag}%
(t+\Delta t)-a%
^{\dag}%
(t) 
\right ] 
&=& 
\left \{ 
\frac{\kappa }{2} 
\left [ 
a%
^{\dag}%
(t) a(t) - 2 
\right ]a%
^{\dag}%
(t) + \mu%
^{\dag} 
(t) 
\right \} \Delta t 
, 
\label{eq:WpB} 
\end{eqnarray}%
where 
\begin{eqnarray} 
\mu (t) &=& \bar\chi(t) \sqrt{p(t)\bar\eta(t)} + 
q%
^{\dag}%
(t) \bar\eta%
^{\dag *}%
(t) 
, 
\\ 
\mu%
^{\dag} 
(t) &=& \bar\chi%
^{\dag}%
(t) \sqrt{p%
^{\dag}%
(t)\bar\eta%
^{\dag}%
(t)} + 
q(t) \bar\eta^{*}(t) 
. 
\end{eqnarray}%
If we require the average random excursion squared at each 
time step, 
\begin{eqnarray} 
\label{eq:Mu2} 
\Delta t^2 
\left [ 
\overline{\hspace{0.1ex} 
\left | 
\mu (t) 
\right | ^2 +  
\left | 
\mu%
^{\dag} 
(t) 
\right | ^2 
\hspace{0.1ex}} 
\right ] 
= \Delta t^2 
\left \{ 
\sqrt{\pi }\, 
\left [  
\left | 
p(t) 
\right | 
+  
\left | 
p%
^{\dag}%
(t) 
\right | 
\right ] 
+  
\left | 
q(t) 
\right |^2 
+  
\left | 
q%
^{\dag}%
(t) 
\right |^2 
\right \} 
, 
\end{eqnarray}%
to be minimal, we find, 
\begin{eqnarray} 
\left | 
q(t) 
\right |^3 = 
\frac{\sqrt{\pi }\,\kappa  
\left | 
a(t) 
\right |  
}{8\Delta t^2} , \ \ \  
\left | 
q%
^{\dag}%
(t) 
\right |^3 = 
\frac{\sqrt{\pi }\,\kappa  
\left | 
a%
^{\dag}%
(t) 
\right |  
}{8\Delta t^2} 
. 
\end{eqnarray}%
Apart from 
Eq.\ (\protect\ref{eq:PQ})%
), 
phases of the $p$'s and $q$'s are of no physical consequence. 
We choose: 
\begin{eqnarray} 
q(t) &=& 
\left [ 
\frac{\sqrt{\pi }\,\kappa 
a(t)  
}{8\Delta t^2} 
\right ]^{1/3} 
, \ \ 
p(t) = - \frac{2}{\sqrt{\pi }} \left [ 
\frac{\sqrt{\pi }\,\kappa 
a(t)  
}{8\Delta t^2} 
\right ]^{2/3}  
, 
\\ 
q%
^{\dag}%
(t) &=& 
\left [ 
\frac{\sqrt{\pi }\,\kappa 
a%
^{\dag}%
(t)  
}{8\Delta t^2} 
\right ]^{1/3} 
, \ \ 
p%
^{\dag}%
(t) = - \frac{2}{\sqrt{\pi }} \left [ 
\frac{\sqrt{\pi }\,\kappa 
a%
^{\dag}%
(t)  
}{8\Delta t^2} 
\right ]^{2/3} 
. 
\end{eqnarray}%
For (\ref{eq:Mu2}), we then find, 
\begin{eqnarray} 
\Delta t^2 
\left [ 
\overline{\hspace{0.1ex} 
\left | 
\mu (t) 
\right | ^2 +  
\left | 
\mu%
^{\dag} 
(t) 
\right | ^2 
\hspace{0.1ex}} 
\right ] \propto \Delta t^{2/3} 
. 
\end{eqnarray}%
This should be compared with the scaling 
characteristic of a Wiener process, $\propto \Delta t$. 
The ``shortage'' of $\Delta t^{1/3}$ results 
in an overall random excursion squared over time $T$ 
scaling as, $\propto T/\Delta t^{1/3}$. 
This is indeed better that the scaling, 
$\propto T/\Delta t^{1/2}$, characteristic of the 
``na\"\i ve'' factorisation, 
Eq.\ (\protect\ref{eq:Naive})%
, yet the improvement is only quantitative, 
the process still diverging in the continuous time limit. 
From a practical perspective, this means that 
sampling noise increases with decreasing time step. 

Another way of factorising the third order terms 
employs three complex HSTs: 
\begin{eqnarray} 
-\frac{\kappa }{8}\xi 
^{\dag 2} 
\xi a 
&%
\stackrel{%
\bar\eta%
}{\Longrightarrow} 
& 
\frac{p\Delta t}{2}\bar\eta \xi 
^{\dag} 
\xi 
+ q \xi%
^{\dag}%
\bar\eta^* , 
\\ 
-\frac{\kappa }{8}\xi ^{2} \xi%
^{\dag} 
a%
^{\dag} 
&%
\stackrel{%
\bar\eta%
^{\dag}%
}{\Longrightarrow} 
& 
\frac{p%
^{\dag}%
\Delta t}{2}\bar\eta%
^{\dag} 
\xi 
^{\dag} 
\xi 
+ q%
^{\dag} 
\xi\bar\eta%
^{\dag *} 
, 
\\ 
\frac{p\bar\eta+p%
^{\dag}%
\bar\eta%
^{\dag}%
}{2} \xi 
^{\dag} 
\xi\Delta t 
&%
\stackrel{%
\bar\eta'%
}{\Longrightarrow} 
& 
\sqrt{p\bar\eta+p%
^{\dag}%
\bar\eta%
^{\dag}%
}\, 
\left ( 
r\bar\eta '\xi%
^{\dag} 
+ 
r%
^{\dag}%
\bar\eta^{\prime *} \xi 
\right ) 
. 
\end{eqnarray}%
As above, the $p$'s and $q$'s here 
are arbitrary functions obeying 
Eq.\ (\protect\ref{eq:PQ})%
, and $r(t)r%
^{\dag}%
(t)=1/2$. 
That is, we recover the generic 
equations (\ref{eq:Wise}) and (\ref{eq:WiseD}), 
this time with 
\begin{eqnarray} 
\mu (t) &=& 
r\bar\eta '\sqrt{p\bar\eta+p%
^{\dag}%
\bar\eta%
^{\dag}%
} 
+ q\bar\eta^* , \\ 
\mu%
^{\dag} 
(t) &=& 
r%
^{\dag}%
\bar\eta^{\prime *}\sqrt{p\bar\eta+p%
^{\dag}%
\bar\eta%
^{\dag}%
} 
+ q%
^{\dag}%
\bar\eta%
^{\dag *} 
. 
\end{eqnarray}%
(Note that the terms proportional to 
the $q$s have exchanged places.) 
Minimising the random excursion squared, 
Eq.\ (\protect\ref{eq:Mu2})%
, 
we find the optimised values of free parameters to be 
\begin{eqnarray} 
r(t) 
&=& 
r%
^{\dag}%
(t)  
= \frac{1}{\sqrt{2}}\, , 
\\ 
q(t) &=& 
\left [ 
\frac{\pi\kappa^2 
a^3(t)  
}{64\Delta t^4 
\left (  
\left | 
a(t) 
\right | 
+  
\left | 
a%
^{\dag}%
(t) 
\right | 
\right ) 
} 
\right ]^{1/6} 
, 
\\ 
q%
^{\dag}%
(t) &=& 
\left [ 
\frac{\pi\kappa^2 
a%
^{\dag 3}%
(t)  
}{64\Delta t^4 
\left (  
\left | 
a(t) 
\right | 
+  
\left | 
a%
^{\dag}%
(t) 
\right | 
\right ) 
} 
\right ]^{1/6} 
, 
\end{eqnarray}%
with the $p$s being recovered from 
Eq.\ (\protect\ref{eq:PQ})%
. 
Evidently both ways of factorising the third 
order terms result in the same scaling of the 
noises. 
\subsection{Numerical experiment} 
We have calculated the time dependence of 
the quadrature amplitude, 
\begin{eqnarray} 
\label{eq:X} 
X(t) = 
\left\langle  
\hat{\sl a}(t)+\hat{\sl a}%
^{\dag}%
(t) 
\right\rangle 
= 
\overline{\hspace{0.1ex} 
a(t)+a%
^{\dag}%
(t) 
\hspace{0.1ex}} 
, 
\end{eqnarray}%
assuming that the initial state of the oscillator is a 
coherent state $%
\left|%
\alpha 
\right\rangle%
$. 
An analytical expression for this quantity may be 
found in \cite{DFW}. 
In 
Fig.\ \protect\ref{figErr}%
, we plot the relative error in $X(t)$, 
\begin{eqnarray} 
\label{eq:DlX} 
\delta X = \frac{X_{\text{simulated}} (t)- 
X_{\text{exact}} (t)}{X_{\text{exact}} (t)} 
, 
\end{eqnarray}%
where $X_{\text{simulated}} (t)$ 
is found via stochastic simulations of the 
``$+$W'' equations, (\ref{eq:WpA})--(\ref{eq:WpB}), 
for $\kappa =4$ and $\alpha =1$. 
To have an idea how important the third order 
noises are, we also calculated $X(t)$ using 
the truncated Wigner representation \cite{Graham}. 
The latter is obtained 
by dropping the noise sources in 
Eqs.\ (\protect\ref{eq:WpA})%
--(\ref{eq:WpB}). 
The resulting equations are readily solved analytically; 
the averaging over the initial W-distribution 
does not pose a problem. 
The relative error in $X(t)$ found via the truncated 
Wigner (``$-$W'') representation is also shown in 
Fig.\ \protect\ref{figErr}%
. 
As is evident from inspection of 
Fig.\ \protect\ref{figErr}%
, 
the ``$+$W'' result is a manifest improvement 
over the ``$-$W'' one. 
The bad news is, however, that the sampling error 
remains rather large, even after averaging over 
more than half a billion trajectories. 
A further piece of bad news is that simulations fell victim 
to numerical instability shortly after the maximal time 
shown in 
Fig.\ \protect\ref{figErr}%
. 
These problems notwithstanding, the very fact that third order 
noises my be subject to stochastic simulations has been clearly 
established. 
\section{The cookbook} 
\label{SecCook} 
\subsection{The general recipe} 
We now summarise our results, recipe-style, the 
way they should be applied to practical calculations. 
For an $n$-mode system, Schr\"odinger-picture 
annihilation and creation operators are 
vectors in respect of the mode index, 
\begin{eqnarray} 
\hat{%
{\mbox{\rm\boldmath$a$}}%
}_{\text{S}} = \left\{ 
\hat{a}_{\text{S}k} 
\right\} , 
\ \ 
\hat{%
{\mbox{\rm\boldmath$a$}}%
}%
_{\text{S}}%
^{\dag} 
= \left\{ 
\hat{a}_{\text{S}k}%
^{\dag} 
\right\} , 
\ \ 
k=1,\cdots,n\, 
, 
\end{eqnarray}%
as are the interaction-picture field operators, 
\begin{eqnarray} 
\hat{%
{\mbox{\rm\boldmath$a$}}%
}(t) = \left\{ 
\hat{a}_k(t) 
\right\} , 
\ \ 
\hat{%
{\mbox{\rm\boldmath$a$}}%
}%
^{\dag}%
(t) = \left\{ 
\hat{a}_k%
^{\dag}%
(t) 
\right\} , 
\ \ 
k=1,\cdots,n\, 
. 
\end{eqnarray}%
The system Hamiltonian consists, as usual, 
of the free and interaction 
Hamiltonians, cf.\ 
Eq.\ (\protect\ref{eq:H})%
. 
The free Hamiltonian is an arbitrary matrix, 
$H = \left\{ 
H_{kk'}\right\} $, $k,k'=1,\cdots,n$, 
in the mode indices. The free 
Schr\"odinger equation thus reads 
\begin{eqnarray} 
i \frac{d\hat{%
{\mbox{\rm\boldmath$a$}}%
}(t)}{dt} = 
H \hat{%
{\mbox{\rm\boldmath$a$}}%
}(t) . 
\end{eqnarray}%
Quantum-classical mappings yield the 
generic system of $2\times n$ 
equations, 
\begin{eqnarray} 
\label{eq:GenA} 
i \frac{d{%
{\mbox{\rm\boldmath$a$}}%
}(t)}{dt} &=& 
H {%
{\mbox{\rm\boldmath$a$}}%
}(t) 
+ 
{\mbox{\rm\boldmath$%
\sigma 
$}}%
(t) 
+ 
{\mbox{\rm\boldmath$%
\sigma 
$}}%
'(t) , 
\\ 
- i \frac{d{%
{\mbox{\rm\boldmath$a$}}%
}%
^{\dag}%
(t)}{dt} &=& 
H {%
{\mbox{\rm\boldmath$a$}}%
^{\dag}%
}(t) 
+ 
{\mbox{\rm\boldmath$%
\sigma 
$}}%
^{\dag}%
(t) 
+ 
{\mbox{\rm\boldmath$%
\sigma 
$}}%
^{\prime\dagger}(t) 
, 
\label{eq:GenB} 
\end{eqnarray}%
for $2\times n$ c-number random fields, 
\begin{eqnarray} 
\label{eq:GenF} 
{%
{\mbox{\rm\boldmath$a$}}%
}(t) = \left\{ 
{a}_k(t) 
\right\} , 
\ \ 
{%
{\mbox{\rm\boldmath$a$}}%
}%
^{\dag}%
(t) = \left\{ 
{a}_k%
^{\dag}%
(t) 
\right\} , 
\ \ 
k=1,\cdots,n\, 
. 
\end{eqnarray}%
The vectors of random sources, 
$%
{\mbox{\rm\boldmath$%
\sigma 
$}}%
'(t),%
{\mbox{\rm\boldmath$%
\sigma 
$}}%
^{\prime\dagger}(t)$, 
depend on the interaction and on the actual type of 
operator ordering underlying the mapping. 

Consider firstly the case of time-normal ordering. 
Without the given sources 
(i.e., with $%
{\mbox{\rm\boldmath$%
\sigma 
$}}%
(t) = 
{\mbox{\rm\boldmath$%
\sigma 
$}}%
^{\dagger}(t) = 0$), 
stochastic averages of the random fields coincide with the 
quantum averages of the time-normally-ordered products of the 
Heisenberg field operators, 
$\hat{\sl a}_k(t),\hat{\sl a}_k%
^{\dag}%
(t)$, $k=1,\cdots,n$. 
With the given sources, 
Eqs.\ (\protect\ref{eq:GenA})%
--(\ref{eq:GenB}) constitute a 
constructive nonlinear-reaction formulation of 
the quantum system. 
The random sources are deduced from the normal form of the 
interaction Hamiltonian, 
\begin{eqnarray} 
\label{eq:Xi3} 
\hat {\cal H}_{\text{int}} = \ 
: 
h\left( 
\hat{%
{\mbox{\rm\boldmath$a$}}%
}_{\text{S}},\hat{%
{\mbox{\rm\boldmath$a$}}%
}%
_{\text{S}}%
^{\dag} 
\right) :\, . 
\end{eqnarray}%
Namely, one should calculate the functional, (cf.\ 
Eq.\ (\protect\ref{eq:Sint})%
) 
\begin{eqnarray} 
S_{\text{int}} \left ( 
{\mbox{\rm\boldmath$\xi$}} 
,%
{\mbox{\rm\boldmath$\xi$}} 
^{\dag}%
,%
{\mbox{\rm\boldmath$a$}} 
,%
{\mbox{\rm\boldmath$a$}} 
^{\dag} 
\right ) 
&=& 
i \int dt 
\left [ 
h 
\left ( 
{\mbox{\rm\boldmath$a$}}%
^{\dag}%
(t) ,%
{\mbox{\rm\boldmath$a$}}%
(t) - i 
{\mbox{\rm\boldmath$\xi$}} 
(t) 
\right ) 
- 
h 
\left ( 
{\mbox{\rm\boldmath$a$}}%
^{\dag}%
(t) + i 
{\mbox{\rm\boldmath$\xi$}}%
^{\dag} 
(t),%
{\mbox{\rm\boldmath$a$}}%
(t) 
\right ) 
\right ] . 
\end{eqnarray}%
Since $\hat {\cal H}_{\text{int}}$ is Hermitian, 
this can be further simplified resulting in 
\begin{eqnarray} 
\nonumber 
S_{\text{int}} \left ( 
{\mbox{\rm\boldmath$\xi$}} 
,%
{\mbox{\rm\boldmath$\xi$}} 
^{\dag}%
,%
{\mbox{\rm\boldmath$a$}} 
,%
{\mbox{\rm\boldmath$a$}} 
^{\dag} 
\right ) 
&=& 
i \int dt\, 
h 
\left ( 
{\mbox{\rm\boldmath$a$}}%
^{\dag}%
(t) ,%
{\mbox{\rm\boldmath$a$}}%
(t) - i 
{\mbox{\rm\boldmath$\xi$}} 
(t) 
\right ) 
+ \,\text{conj.} , 
\end{eqnarray}%
where conjugation acts as a formal Hermitian 
transformation (i.e., it interchanges quantities with 
and without dagger, 
$%
{\mbox{\rm\boldmath$a$}}%
(t)\leftrightarrow 
{\mbox{\rm\boldmath$a$}}%
^{\dag}%
(t)$, 
$%
{\mbox{\rm\boldmath$\xi$}}%
(t)\leftrightarrow 
{\mbox{\rm\boldmath$\xi$}}%
^{\dag}%
(t)$, 
and complex-conjugates other c-numbers). 
One should then factorise powers and products of 
$\xi $s, employing a suitable set of Hubbard-Stratonovich 
transformations (see 
Eqs.\ (\protect\ref{eq:%
HSTArr%
})%
), 
until an expression 
emerges which is already 
linear in respect of the $\xi $s: 
\begin{eqnarray} 
S_{\text{int}} \left ( 
{\mbox{\rm\boldmath$\xi$}} 
,%
{\mbox{\rm\boldmath$\xi$}} 
^{\dag}%
,%
{\mbox{\rm\boldmath$a$}} 
,%
{\mbox{\rm\boldmath$a$}} 
^{\dag} 
\right ) 
\ 
\cdots 
\stackrel{%
\cdots%
}{\longrightarrow}%
\cdots%
\stackrel{%
\cdots%
}{\Longrightarrow}%
\cdots 
\ 
\int dt \sum_{k=1}^n 
\left[ 
\xi _k (t) \sigma _k^{\prime\dagger} (t) + 
\xi _k%
^{\dag} 
(t) \sigma _k' (t) 
\right] . 
\end{eqnarray}%
(Note that terms without $\xi$s in 
$S_{\text{int}}$ always cancel.) 
The $\sigma $s thus obtained are exactly those which appear in 
Eqs.\ (\protect\ref{eq:GenA})%
--(\ref{eq:GenB}). 
If $S_{\text{int}} $ is at most quadratic in $\xi $'s, 
Eqs.\ (\protect\ref{eq:GenA})%
--(\ref{eq:GenB}) are a system of genuine 
It\^o stochastic differential equations. 
Otherwise they can only be interpreted as 
difference equations 
over a finite time step $\Delta t$. 

In the case of the time-Wigner ordering, this recipe 
applies with the replacement 
$S_{\text{int}}\to S^W_{\text{int}}$, 
where 
\begin{eqnarray} 
\nonumber 
S^W_{\text{int}} \left ( 
{\mbox{\rm\boldmath$\xi$}} 
,%
{\mbox{\rm\boldmath$\xi$}} 
^{\dag}%
,%
{\mbox{\rm\boldmath$a$}} 
,%
{\mbox{\rm\boldmath$a$}} 
^{\dag} 
\right ) 
&=& i \int dt 
\left [ 
h^W 
\left ( 
{\mbox{\rm\boldmath$a$}}%
^{\dag}%
(t) - \frac{i}{2} 
{\mbox{\rm\boldmath$\xi$}}%
^{\dag} 
(t), 
{\mbox{\rm\boldmath$a$}}%
(t) - \frac{i}{2} 
{\mbox{\rm\boldmath$\xi$}} 
(t) 
\right )\right. 
\nonumber 
\\ 
&& -\, \left. 
h^W 
\left ( 
{\mbox{\rm\boldmath$a$}}%
^{\dag}%
(t) + \frac{i}{2} 
{\mbox{\rm\boldmath$\xi$}}%
^{\dag} 
(t), 
{\mbox{\rm\boldmath$a$}}%
(t) + \frac{i}{2} 
{\mbox{\rm\boldmath$\xi$}} 
(t) 
\right ) 
\right ] 
\\ 
&=& i \int dt\, 
h^W 
\left ( 
{\mbox{\rm\boldmath$a$}}%
^{\dag}%
(t) - \frac{i}{2} 
{\mbox{\rm\boldmath$\xi$}}%
^{\dag} 
(t), 
{\mbox{\rm\boldmath$a$}}%
(t) - \frac{i}{2} 
{\mbox{\rm\boldmath$\xi$}} 
(t) 
\right ) 
+ \, \text{conj.}, 
\end{eqnarray}%
and $h^W$ is the symmetric form of the 
interaction Hamiltonian, 
\begin{eqnarray} 
\label{eq:XiW3} 
\hat {\cal H}_{\text{int}} = 
\text{W} 
\left\{ h^W 
\left( 
\hat{%
{\mbox{\rm\boldmath$a$}}%
}_{\text{S}},\hat{%
{\mbox{\rm\boldmath$a$}}%
}%
_{\text{S}}%
^{\dag} 
\right) 
\right\} 
. 
\end{eqnarray}%
We are however not aware of any nonlinear system 
where $S^W_{\text{int}}$ would turn out to be quadratic 
in the $\xi $'s. 
\subsection{Degenerate OPO} 
\subsubsection{Generalised positive-P representation}\label{GenOPO} 
We shall now 
illustrate the general recipe by deriving 
the positive-P and positive-W representations 
for a degenerate optical 
parametric oscillator (OPO). 
The degenerate OPO 
consists of two coupled 
oscillators described by the Hamiltonian, 
\begin{eqnarray} 
{\cal H} = 
\omega \hat a_{\text{S}1} \hat a_{\text{S}1}%
^{\dag} 
+ 
2\omega \hat a_{\text{S}2} \hat a_{\text{S}2}%
^{\dag} 
+ 
\frac{i\kappa }{2} 
\left[ 
\hat a_{\text{S}1}%
^{\dag 2}%
\hat a_{\text{S}2} 
- 
\hat a_{\text{S}1}^{2}\hat a_{\text{S}2}%
^{\dag} 
\right]. 
\end{eqnarray}%
Then, in brief notation, 
\begin{eqnarray} 
\nonumber 
S_{\text{int}} \left( 
{\mbox{\rm\boldmath$\xi$}} 
,%
{\mbox{\rm\boldmath$\xi$}} 
^{\dag}%
,%
{\mbox{\rm\boldmath$a$}} 
,%
{\mbox{\rm\boldmath$a$}} 
^{\dag} 
\right) =&& i \frac{i\kappa }{2} 
\left[ 
a_{1}%
^{\dag 2} 
\left( a_{2} - i\xi_2 \right) 
- 
\left( a_{1} - i \xi_1 \right)^{2} 
a_{2}%
^{\dag} 
\right] 
+ \,\text{conj.} 
\\ 
=&& 
-i\kappa\xi_1 a_1 a_2%
^{\dag} 
+ i \frac{\kappa}{2}\xi_2 a_1%
^{\dag 2} 
- \frac{\kappa}{2}\xi_1^2 a_2%
^{\dag} 
+ \,\text{conj.} 
\nonumber 
\\ 
\stackrel{%
\chi,\chi%
^{\dag}%
}{\longrightarrow} 
&& 
+i \xi_1%
^{\dag} 
\left( 
\kappa a_1%
^{\dag} 
a_2 + \chi\sqrt{\kappa a_2} 
\right) 
\nonumber 
\\ 
&& 
-i \xi_1 \left( 
\kappa a_1 a_2%
^{\dag} 
+ \chi%
^{\dag}%
\sqrt{\kappa a_2%
^{\dag}%
} 
\right) 
\nonumber 
\\ 
&& 
- i \xi_2%
^{\dag} 
\frac{\kappa}{2} a_1^{2} 
\nonumber 
\\ 
&& 
+ i \xi_2 \frac{\kappa}{2} a_1%
^{\dag 2} 
, 
\end{eqnarray}%
where $\chi(t),\chi%
^{\dag}%
(t)$ are a pair 
of independent real $\delta$-correlated Gaussian 
noises. 
The positive-P representation 
for the OPO, generalised to the response problem, 
then reads, 
\begin{eqnarray} 
\frac{da_1(t)}{dt} &=& 
- i \sigma_1(t) + 
\kappa a_1%
^{\dag}%
(t) a_2(t) 
+ \chi(t)\sqrt{\kappa a_2(t)}, 
\label{eq:OPOA} 
\\ 
\frac{da_1%
^{\dag}%
(t)}{dt} &=& 
+i \sigma_1%
^{\dag}%
(t) + 
\kappa a_1(t) a_2%
^{\dag}%
(t) 
+ \chi%
^{\dag}%
(t)\sqrt{\kappa a_2%
^{\dag}%
(t)}, 
\\ 
\frac{da_2(t)}{dt} &=& 
- i \sigma_2(t) - \frac{\kappa}{2} a_1^{2}(t) , 
\\ 
\label{eq:OPOD} 
\frac{da_2%
^{\dag}%
(t)}{dt} &=& 
+ i \sigma_2%
^{\dag}%
(t) - \frac{\kappa}{2} a_1%
^{\dag 2}%
(t) 
. 
\end{eqnarray}%
This derivation is strikingly simple and 
straightforward, and compares very favorably 
to the common derivation 
based on phase-space techniques \cite{DFW} 
(not to mention that 
Eqs.\ (\protect\ref{eq:OPOA})%
--(\ref{eq:OPOD}) 
allow for a much deeper insight). 
\subsubsection{Positive-W representation} 
Similarly, 
\begin{eqnarray} 
\nonumber 
S^W_{\text{int}} \left( 
{\mbox{\rm\boldmath$\xi$}} 
,%
{\mbox{\rm\boldmath$\xi$}} 
^{\dag}%
,%
{\mbox{\rm\boldmath$a$}} 
,%
{\mbox{\rm\boldmath$a$}} 
^{\dag} 
\right) =&& i \frac{i\kappa }{2} 
\left[ 
\left( a_{1}%
^{\dag} 
- i\xi_1%
^{\dag}%
/2 \right)^2 
\left( a_{2} - i\xi_2/2 \right) 
- 
\left( a_{1} - i \xi_1/2 \right)^{2} 
\left( a_{2}%
^{\dag} 
- i\xi_2%
^{\dag}%
/2 \right) 
\right] 
+ \,\text{conj.} 
\\ 
=&& 
-i\kappa\xi_1 a_1 a_2%
^{\dag} 
+ i \frac{\kappa}{2}\xi_2 a_1%
^{\dag 2} 
+ \frac{i\kappa}{8}\xi_1^2 \xi_2%
^{\dag} 
+ \,\text{conj.} 
\nonumber 
\\ 
\stackrel{%
\eta ,\eta 
^{\dag}%
}{\Longrightarrow} 
&& 
-i\kappa\xi_1 a_1 a_2%
^{\dag} 
+ i \frac{\kappa}{2}\xi_2 a_1%
^{\dag 2} 
- \frac{ p%
^{\dag}%
}{2}\eta ^*\xi_1^2 
+ iq\eta \xi_2%
^{\dag} 
+ \,\text{conj.} 
\nonumber 
\\ 
\stackrel{%
\chi,\chi%
^{\dag}%
}{\longrightarrow} 
&& 
+i \xi_1%
^{\dag} 
\left( 
\kappa a_1%
^{\dag} 
a_2 + \chi\sqrt{p\eta 
^{\dag *}%
} 
\right) 
\nonumber 
\\ 
&& 
-i \xi_1 \left( 
\kappa a_1 a_2%
^{\dag} 
+ \chi%
^{\dag}%
\sqrt{p 
^{\dag}%
\eta^{*}} 
\right) 
\nonumber 
\\ 
&& 
+ i \xi_2%
^{\dag} 
\left( 
- \frac{\kappa}{2} a_1^{2} +q\eta 
\right) 
\nonumber 
\\ 
&& 
- i \xi_2 \left( 
- \frac{\kappa}{2} a_1%
^{\dag 2} 
+ q%
^{\dag} 
\eta 
^{\dag} 
\right) 
, 
\end{eqnarray}%
resulting in the positive-W representation 
for the OPO: 
\begin{eqnarray} 
\frac{da_1(t)}{dt} &=& 
- i \sigma_1(t) + 
\kappa a_1%
^{\dag}%
(t) a_2(t) 
+ \chi(t)\sqrt{p(t)\eta 
^{\dag *}%
(t)}, 
\label{eq:OPOWA} 
\\ 
\frac{da_1%
^{\dag}%
(t)}{dt} &=& 
+i \sigma_1%
^{\dag}%
(t) + 
\kappa a_1(t) a_2%
^{\dag}%
(t) 
+ \chi%
^{\dag}%
(t)\sqrt{p 
^{\dag}%
(t)\eta^{*}(t)}, 
\\ 
\frac{da_2(t)}{dt} &=& 
- i \sigma_2(t) - \frac{\kappa}{2} a_1^{2}(t) 
+ q(t) \eta (t) 
, 
\\ 
\label{eq:OPOWD} 
\frac{da_2%
^{\dag}%
(t)}{dt} &=& 
+ i \sigma_2%
^{\dag}%
(t) - \frac{\kappa}{2} a_1%
^{\dag 2}%
(t) 
+ q%
^{\dag}%
(t) \eta%
^{\dag} 
(t) 
. 
\end{eqnarray}%
Here, 
$\eta (t),\eta 
^{\dag}%
(t)$ and $\chi (t),\chi 
^{\dag}%
(t)$ are 
pairs of standardised Gaussian noises (respectively, 
complex and real ones), 
and 
the functions $p(t),p%
^{\dag}%
(t),q(t),q%
^{\dag}%
(t)$ obey the 
relations, 
\begin{eqnarray} 
\label{eq:PQOPO} 
p(t)q%
^{\dag}%
(t) = p%
^{\dag}%
(t)q(t) = -\frac{\kappa }{4} . 
\end{eqnarray}%
Equations (\ref{eq:OPOWA})--(\ref{eq:OPOWD}) are 
difference equations over a {\em finite\/} time step 
$d t = \Delta t$, 
with the noises normalised such that 
\begin{eqnarray} 
\overline{\hspace{0.1ex}%
\chi (t)\chi (t')%
\hspace{0.1ex}} 
= 
\overline{\hspace{0.1ex}%
\chi%
^{\dag} 
(t)\chi%
^{\dag} 
(t')%
\hspace{0.1ex}} 
= 
\overline{\hspace{0.1ex}%
\eta (t)\eta^* (t')%
\hspace{0.1ex}} 
= 
\overline{\hspace{0.1ex}%
\eta%
^{\dag} 
(t)\eta%
^{\dag *} 
(t')%
\hspace{0.1ex}} 
= \delta (t-t') 
= \frac{\delta _{tt'}}{\sqrt{dt}} 
. 
\end{eqnarray}%
The $p$'s and $q$'s will be chosen so as to minimise the 
weighted sum of random increments squared, 
\begin{eqnarray} 
\nonumber 
&\displaystyle dt^2\left\{ 
\overline{\hspace{0.1ex} 
\left | 
\chi(t)\sqrt{p(t)\eta 
^{\dag *}%
(t)} 
\right | ^2 
\hspace{0.1ex}} 
+ 
\overline{\hspace{0.1ex} 
\left | 
\chi%
^{\dag}%
(t)\sqrt{p 
^{\dag}%
(t)\eta^{*}(t)} 
\right | ^2 
\hspace{0.1ex}} 
+ w\left[ 
\overline{\hspace{0.1ex} 
\left | 
q(t) \eta (t) 
\right | ^2 
\hspace{0.1ex}} 
+ 
\overline{\hspace{0.1ex} 
\left | 
q%
^{\dag}%
(t) \eta%
^{\dag} 
(t) 
\right | ^2 
\hspace{0.1ex}} 
\right] 
\right\} \\ 
&\displaystyle \hspace{5ex}= 
{p(t)\sqrt{\pi dt}} + 
\frac{w\kappa ^2 dt}{16p^2(t)} 
+ \,\text{conj.} 
, 
\label{eq:Inc2W} 
\end{eqnarray}%
where we have chosen the 
$p$'s to be real and positive. 
The weighting factor $w>0$ allows one to redistribute the noise 
between the fundamental and the subharmonic, and may 
be fine-tuned by trial and error when running stochastic 
simulations. 
Minimising (\ref{eq:Inc2W}) yields: 
\begin{eqnarray} 
\label{eq:POPO} 
p(t) = p%
^{\dag}%
(t) = 
\frac{1}{2}\sqrt[3]{w\kappa^2 \sqrt{{dt}/{\pi } }}\, . 
\end{eqnarray}%
The $q$'s are then recovered from (\ref{eq:PQOPO}). 
Accounting for (\ref{eq:POPO}), all noise increments 
in the positive-W equations scale as, $\propto d t^{1/3}$. 
 
\begin{figure} 
\begin{center} 
\setlength{\unitlength}{0.6\textwidth} 
\begin{picture}(1,0.4)(0,0.0) 
\put(0,0.13){\line(1,0){0.95}} 
\put(0,0.23){\line(1,0){0.95}} 
\put(0.95,0.18){\oval(0.1,0.1)[r]} 
\put(0.0,0.17){$t = -\infty$} 
\put(0.80,0.17){$t = +\infty$} 
\put(0.1,0.13){\vector(-1,0){0}} 
\put(0.1,0.23){\vector(1,0){0}} 
\put(1,0.18){\vector(0,-1){0}} 
\thicklines 
\put(0.4,0.23){\oval(0.2,0.15)[t]} 
\put(0.41,0.305){\vector(1,0){0}} 
\put(0.5,0.23){\circle*{0.01}} 
\put(0.3,0.23){\circle*{0.01}} 
\put(0.205,0.3){$iG_{++}$} 
\put(0.3,0.13){\oval(0.2,0.15)[b]} 
\put(0.29,0.055){\vector(-1,0){0}} 
\put(0.4,0.13){\circle*{0.01}} 
\put(0.2,0.13){\circle*{0.01}} 
\put(0.4,0.04){$iG_{--}$} 
\put(0.59,0.16){$iG_{-+}$} 
\put(0.71,0.23){\line(0,-1){0.1}} 
\put(0.71,0.17){\vector(0,-1){0}} 
\put(0.71,0.23){\circle*{0.01}} 
\put(0.71,0.13){\circle*{0.01}} 
\end{picture} 
\end{center} 
\caption{The Schwinger-Perel-Keldysh C-contour 
(thin lines) and the 
three contractions (thick lines) contributing to $\Delta_C$, cf.\ 
Eq.\ (\protect\ref{eq:DeltaC}). 
The arrows on contractions are from creation to annihilation, 
cf.\ Eq.\ (\protect\ref{eq:Gab}), 
and the time order of ends corresponds to contractions 
being non-zero.} 
\label{figC} 
\end{figure}
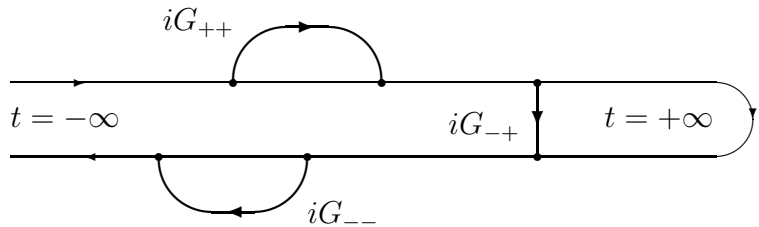 
\begin{figure} 
\begin{center} 
\epsfxsize=0.75\textwidth 
\epsfbox{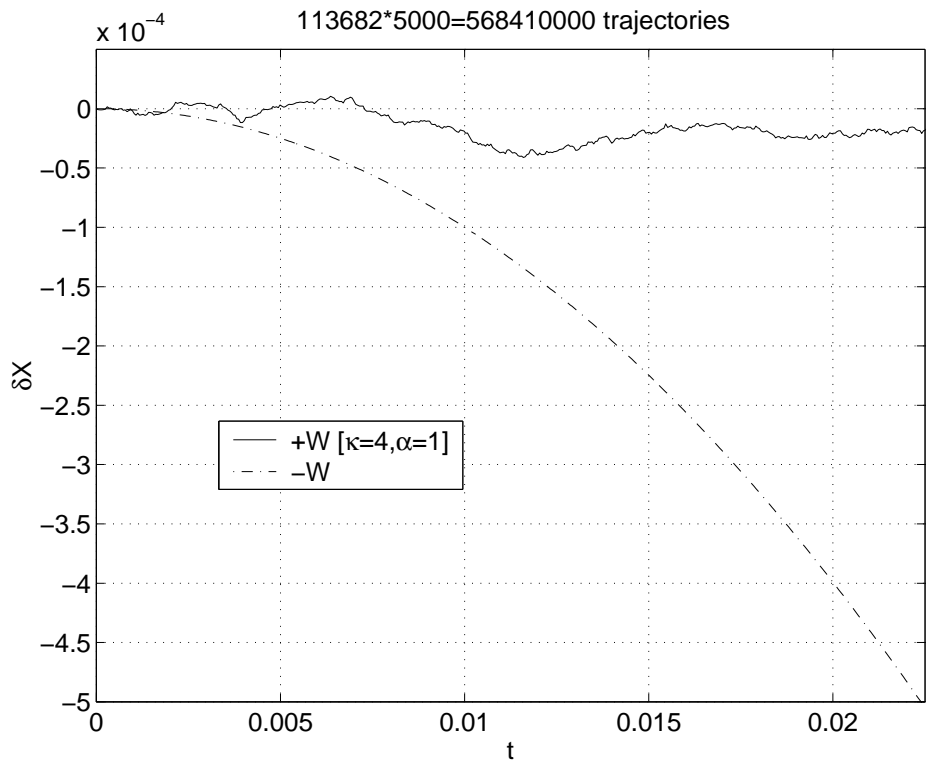} 
\end{center} 
\caption{Relative error in the 
coherent field amplitude $X(t)$, 
calculated using the positive-W ($+$W, solid line) 
and truncated 
Wigner ($-$W, dash-dotted line) representations, 
for nonlinearity $\kappa =4$, and the inital 
coherent state of the oscillator with 
amplitude $\alpha=1$. 
} 
\label{figErr} 
\end{figure} 
\end{document}